\documentclass{article}

\usepackage{arxiv}

\usepackage[utf8]{inputenc} 
\usepackage[T1]{fontenc}    
\usepackage{hyperref}       
\usepackage{url}            
\usepackage{booktabs}       
\usepackage{amsfonts}       
\usepackage{nicefrac}       
\usepackage{microtype}      
\usepackage{graphicx}
\usepackage{doi}

\usepackage[numbers,sort&compress]{natbib}

\usepackage{caption}
\usepackage[labelformat=simple]{subcaption}

\usepackage{graphicx}
\usepackage{textcomp}
\usepackage{amsmath}
\usepackage{float}
\usepackage{siunitx}
\usepackage{array}
\usepackage{booktabs}
\usepackage{longtable}
\usepackage{multirow}
\usepackage{multicol}

\newcommand\HLBM{HLBM}
\newcommand\Rey{\mbox{\textit{Re}}}

\usepackage[capitalise]{cleveref}

\title{Potential for damage to fruits during transport through cross-section constrictions}

\author{
	\href{https://orcid.org/0000-0002-7666-3439}{\includegraphics[scale=0.06]{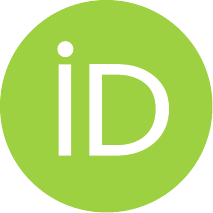}\hspace{1mm}Jan E. Marquardt}\thanks{Corresponding author}\\
	Lattice Boltzmann Research Group \\
	Institute of Mechanical Process Engineering and Mechanics\\
	Karlsruhe Institute of Technology\\
	76131 Karlsruhe, Germany \\
	\texttt{jan.marquardt@kit.edu} \\
	\And
	\href{https://orcid.org/0009-0003-7714-909X}{\includegraphics[scale=0.06]{orcid.pdf}\hspace{1mm}Bastian Eysel}\\
	Institute of Food Biotechnology and Food Process Engineering \\
	Faculty III Process Sciences \\
	Technische Universität Berlin \\
	10623 Berlin, Germany \\
	\And
	\href{https://orcid.org/0009-0000-5678-0390}{\includegraphics[scale=0.06]{orcid.pdf}\hspace{1mm}Martin Sadric}\\
	Lattice Boltzmann Research Group \\
	Institute for Applied and Numerical Mathematics \\
	Karlsruhe Institute of Technology \\
	76131 Karlsruhe, Germany \\
	\And
	\href{https://orcid.org/0000-0002-4696-1128}{\includegraphics[scale=0.06]{orcid.pdf}\hspace{1mm}Cornelia Rauh}\\
	Institute of Food Biotechnology and Food Process Engineering \\
	Faculty III Process Sciences \\
	Technische Universität Berlin \\
	10623 Berlin, Germany \\
	\And
	\href{https://orcid.org/0000-0003-1026-6462}{\includegraphics[scale=0.06]{orcid.pdf}\hspace{1mm}Mathias J. Krause}\\
	Lattice Boltzmann Research Group \\
	Institute for Applied and Numerical Mathematics \\
	Karlsruhe Institute of Technology \\
	76131 Karlsruhe, Germany \\
}

\renewcommand{\headeright}{J.\ E.\ Marquardt \emph{et al.}}
\renewcommand{\shorttitle}{Potential for damage to fruits}

\hypersetup{
	pdftitle={FruitDamagePotential},
	pdfsubject={},
	pdfauthor={Jan E. Marquardt}
}

\begin{document}
	\maketitle
	
\begin{abstract}
	Fruit preparations are used in various forms in the food industry. For example, they are used as an ingredient in dairy products such as yogurt with added fruit. The dispersed fruit pieces can be described as soft particles with viscoelastic material behavior. The continuous phase is represented by fluids with complex flow behavior depending on the formulation. Characterization has shown that this can be described by the Herschel--Bulkley model. Since damage to fruit pieces is undesirable in industrial transport processes, the potential for damage to fruit pieces during transport of pipes in cross-sectional constrictions is analyzed. The analysis is performed numerically using the homogenized lattice Boltzmann method and validated by an experiment on industrial fruit preparations at pilot plant scale. The results show a strong dependence of the damage potential on the (local) Metzner--Reed Reynolds number.
\end{abstract}


\section{Introduction}
\label{sec:intro}
Many food products can be described as suspensions of mechanically sensitive, i.e. deformable or fragile, particles.
Examples include chunky fruit preparations commonly used in the dairy industry, such as fruit yogurt.
During the manufacturing process, transport through piping and various fittings (e.g. pipe constrictions/expansions, valves, pumps, flaps, bends) plays a significant role and leads to a high potential for damage.
During this transport, the particle component is subjected to mechanical stresses due to hydrodynamic forces and collisions with other particles or with apparatus parts.
These stresses are highly variable in intensity and transient in nature.
\par
Literature primarily focuses on pipeline transport of coarse-dispersed food suspensions to avoid under-processing during thermal treatment~\cite{barigou2003concentric,legrand2007characterization}.
Flow conditions are characterized based on the solid volume fraction and other dimensionless parameters (Reynolds number, particle Reynolds number, particle Froude number, Stokes number)~\cite{lareo1997fluid}, facilitating the identification of various regimes describing the distribution of the particle component across the pipe~\cite{michaelides2016multiphase,brennen2005fundamentals}.
Such concentrated suspensions must flow at relatively high velocities to maintain a homogeneous particle distribution.
Heterogeneous flow occurs when a gradient in particle concentration is evident, with cases of pronounced gradients distinguished as moving bed and stationary bed formations~\cite{li2017mechanical}.
\par
Laminar flows of suspensions of (usually spherical) particles of similar density to the surrounding fluid have been studied experimentally~\cite{barigou2003concentric,legrand2007characterization,pervez1994long,choi2010inertial,mathai2016translational} and numerically~\cite{chun2006inertial,loisel2013effect,wu2011fully,ardekani2018numerical,eesa2009cfd}.
However, the focus of these works is on the particle-fluid interaction, not the potential damage.
\par
Thus, while the behavior of laminar, particle-laden flows in pipelines is partially understood, this is not the case for mechanical damage to the particulate phase.
Considerations of damage have been limited to harvesting, packaging, and vehicle transport of fresh fruits~\cite{Li2014}.
In the context of food processing, only the influence of pumps on particle damage has been studied using the example of rotary lobe and centrifugal pumps~\cite{Schmitt2018}.
However, unlike pumps, the stress in piping systems can be expected to be cyclical and increasing/decreasing as particles move continuously with the flow.
Individual events are not necessarily binary.
Rather, fruit preparations are subjected to several successive transport processes, so that the entire history plays a role.
The resulting strain is quantified on the basis of collision, shear and elongation effects~\cite{wollny2007partikelbeanspruchung,wille2001pda}.
Their importance for the momentum and energy coupling between disperse and continuous phase depends strongly on the characteristic length scales and the material consistency of the coarsely dispersed phase.
In the case of (approximately) deformation-resistant fruit pieces, the coupling is largely achieved by viscous forces at the surface and normal forces caused by the pressure distribution.
\par
Despite the existing research on laminar particle-laden flows and particle-fluid interactions, there is a significant gap in the understanding of mechanical damage to the particle component, particularly in the context of the transport of food suspensions through piping systems.
The unique stress conditions in such systems have not been investigated, highlighting the need for a comprehensive study to address these knowledge gaps, with a particular focus on the impact of pipeline transport on the quality and integrity of food suspensions.
\par
Therefore, the aim of this research paper is to investigate the damage potential of the particulate component of food suspensions during their flow through piping systems, in particular cross-sectional constrictions, under practical and relevant conditions, including process parameters and material properties.
The objective of the work is to investigate and understand the mutual influence between the continuous and dispersed phases, to identify and analyze the types and intensities of stresses that occur during flow, and to determine which of these stresses are primarily responsible for significant particle damage.
In addition, the spatial dependencies of the flow-induced stresses caused by the geometries of the equipment are investigated, and the main parameters influencing the damage of the particle phase are determined, while the interactions between them are studied.
To achieve these objectives, the current work uses a combination of experiments and simulations, providing a comprehensive and in-depth analysis of the mechanical stresses and damage experienced by food particles during transport through pipes with cross-sectional constrictions.
Ultimately, the knowledge gained supports the design of piping systems in food processing to improve the quality and integrity of food products.
\par
To this end, the following sections are organized as follows.
Section~\ref{sec:modeling} presents the models that describe the fluid and the particle, while Section~\ref{sec:methods} discusses the numerical methods used for the simulations.
In Section~\ref{sec:setups} we outline the experimental and computational setups and in Section~\ref{sec:results} we present the results of the following studies.
Finally, the key findings and conclusions of this study are summarized in Section~\ref{sec:summary}.

\section{Modeling}
\label{sec:modeling}

\subsection{Fluid}
\label{sec:modeling:fluid}
Fluids are generally considered to be incompressible.
In this case, the Navier--Stokes equations become
\begin{align}
	\begin{split}
	\frac{\partial \boldsymbol{u}_\text{f}}{\partial t} + \left( \boldsymbol{u}_\text{f} \cdot \nabla \right) \boldsymbol{u}_\text{f} - \nu \Delta \boldsymbol{u}_\text{f} + \frac{1}{\rho_\text{f}} \nabla p &= \frac{\boldsymbol{F}_\text{f}}{\rho_\text{f}},
	\\
	\nabla \cdot \boldsymbol{u}_\text{f} &= 0,
	\end{split}
\end{align}
where $p$ is the pressure, $t$ is the time, $\boldsymbol{F}_\text{f}$ is the total of all forces acting on the fluid and $\boldsymbol{u}_\text{f}$, $\rho_\text{f}$, $\eta$ are the velocity, density and dynamic viscosity of the fluid.
\par
For Newtonian fluids, $\eta$ is a constant.
For non-Newtonian fluids, however, the viscosity depends on the shear rate $\dot{\gamma}$.
Power laws are usually used to represent this dependency.
Using the Herschel--Bulkley relation, the effective viscosity is given by \cite{Sahu_Valluri_Spelt_Matar_2007}
\begin{align}
    \label{eq:hbviscmodel}
    \eta^* = \tau_0 |\dot{\gamma}|^{-1} + k |\dot{\gamma}|^{n-1},
\end{align}
with the yield stress $\tau_0$, consistency index $k$, and power law index $n$.
Since this equation diverges for $\dot{\gamma} \to 0$, a minimal shear rate $\dot{\gamma}_\text{min}$ is usually used in the equation above for $\dot{\gamma} < \dot{\gamma}_\text{min}$.
\par
When the regarded fluid has a yield stress $\tau_0 = 0$, then the above simplifies to the Ostwald--de Waele relation, for which the effective viscosity reads \cite{Irgens_2014}
\begin{align}
    \label{eq:owviscmodel}
    \eta^* = k |\dot{\gamma}|^{n-1}.
\end{align}

\subsection{Particle}
\label{sec:modeling:particle}

For the particle component, we use Newton's second law of motion.
Therefore, translation is described by
\begin{align}
    \label{eq:particlemotiontrans}
	{m_\text{p}} \frac{\partial \boldsymbol{u}_\text{p}}{\partial t} = \boldsymbol{F}_\text{p}
\end{align}
and rotation by
\begin{equation}
	\label{eq:motionrot}
	\boldsymbol{I}_\text{p} \frac{\partial \boldsymbol{\omega}_\text{p}}{\partial t} +  \boldsymbol{\omega}_\text{p} \times (\boldsymbol{I}_\text{p} \cdot \boldsymbol{\omega}_\text{p}) = \boldsymbol{T}_\text{p}.
\end{equation}
Here, ${m_\text{p}}$, $\boldsymbol{I}_\text{p}$, $\boldsymbol{u}_\text{p}$,  $\boldsymbol{\omega}_\text{p}$ are the mass, moment of inertia, velocity, and angular velocity of the particle, respectively.
$\boldsymbol{F}_\text{p}$ and $\boldsymbol{T}_\text{p}$ are the sum of the acting forces and torque affecting the particle motion, respectively.
Above, the subscript $\text{p}$ indicates that the quantities refer to the particle's center of mass.

\subsection{Contact}
\label{sec:modeling:contact}

In order to accurately model interactions involving complex particle geometries, we employ a contact model that allows for arbitrarily shaped bodies~\cite{Nassauer_Kuna_2013}.
The normal contact force in this model is given by
\begin{equation}
\label{eq:contact_force_normal}
\boldsymbol{F}_{\text{c},\text{n}} = \frac{4}{3 \pi} \boldsymbol{n}_\text{c} E^* \sqrt{V_\text{c} \delta} \left( 1 + c \dot{\delta}_\text{n} \right).
\end{equation}
In this equation, $E^*$ is the effective Young's modulus, $V_\text{c}$ is the overlap volume, $\delta$ is the indentation depth, $c$ is a damping factor, and $\dot{\delta}_\text{n}$ is the relative velocity between two particles in contact along the contact normal $\boldsymbol{n}_\text{c}$.
The effective Young's modulus is
\begin{equation}
E^* = \left( \frac{1 - \nu_\text{A}^2}{E_\text{A}} + \frac{1-\nu^2_\text{B}}{E_\text{B}} \right)^{-1},
\end{equation}
where $E_\text{A}$ and $\nu_\text{A}$ are the Young's modulus and Poisson's ratio of particle A, and $E_\text{B}$ and $\nu_\text{B}$ are those of another solid object B. 
The damping factor $c$ is related to the coefficient of restitution $e$ as follows~\cite{Carvalho_Martins_2019}
\begin{equation}
\label{eq:damping_factor}
c =
\begin{cases}
1.5 \frac{(1-e)(11-e)}{(1+9e)u_0} &\text{for } u_0 > 0 \\
0 &\text{for } u_0 \leq 0
\end{cases},
\end{equation}
where $u_0$ is the relative velocity at the initial contact.
Tangential forces, which are influenced by normal forces and the coefficients of static and kinetic friction ($\mu_\text{s}$ and $\mu_\text{k}$), are described by
\begin{equation}
\label{eq:contact_force_tangential}
\boldsymbol{F}_{\text{c},\text{t}} =
- \frac{\boldsymbol{u}_{\text{AB},\text{t}} (\boldsymbol{x}_\text{c})}{||\boldsymbol{u}_{\text{AB},\text{t}} (\boldsymbol{x}_\text{c})||}
\left( \left( 2 \mu_\text{s}^* - \mu_\text{k} \right) \frac{a^2}{a^4+1} + \mu_\text{k} - \frac{\mu_\text{k}}{a^2+1} \right) ||\boldsymbol{F}_{\text{c},\text{n}}||,
\end{equation}
with
\begin{equation}
\mu_\text{s}^* = \mu_\text{s} \left( 1 - 0.09 \left( \frac{\mu_\text{k}}{\mu_\text{s}} \right)^4 \right),
\end{equation}
and
\begin{equation}
a = \frac{||\boldsymbol{u}_{\text{AB},\text{t}} (\boldsymbol{x}_\text{c})||}{u_k}.
\end{equation}
Here, $\boldsymbol{u}_{\text{AB},\text{t}}$ is the relative tangential between the two solids A and B, and $u_k$ is the velocity at which the transition from static to kinetic friction occurs, which we set to $0.01~\si{\meter\per\second}$.

\section{Numerical Methods}
\label{sec:methods}

In the following, the numerical methods used for the simulations are introduced.
Note that the all values in Section~\ref{sec:methods} are given in lattice units, unless explicitly stated otherwise.
\par
All methods presented below are implemented in the open source software OpenLB~\cite{olb17,Krause2020}, which is used in this work.

\subsection{Lattice Boltzmann method}
\label{sec:methods:lbm}

We use the lattice Boltzmann method (LBM) \cite{Succi_2001,Krueger2016,sukop2006LatticeBoltzmannModeling} to solve the Navier--Stokes equations for incompressible flows. 
\par
LBM has its roots in gas kinetics, which explains why it operates at the mesoscopic level and considers the behavior of fluid particle populations.
The discrete velocity distribution function $f_i(\boldsymbol{x},t)$ is used to characterize these populations at a position $\boldsymbol{x}$ and time $t$.
The index $i$ refers to the corresponding discrete velocity $\boldsymbol{c}_i$, which is given by the selected velocity set.
There are several velocity sets available in literature \cite{Succi_2001,Krueger2016}.
We choose the D$3$Q$19$, which discretizes the three-dimensional space and contains of $19$ discrete velocities, for the studies in this paper.
The populations are also used to derive macroscopic quantities such as density $\rho_\text{f}(\boldsymbol{x},t) = \sum_i f_i(\boldsymbol{x},t)$ and velocity $\rho_\text{f} \boldsymbol{u}_\text{f}(\boldsymbol{x},t) = \sum_i\boldsymbol{c}_i f_i(\boldsymbol{x},t)$.
\par
The particle populations' evolution over time is expressed by the lattice Boltzmann equation that is usually divided into a collision and streaming step.
The former reads
\begin{equation}
    \label{eq:collide}
    f_i^*(\boldsymbol{x}, t) = f_i(\boldsymbol{x}, t) + \Omega_i(\boldsymbol{x}, t) + S_i(\boldsymbol{x}, t).
\end{equation}
Here, the post-collision distribution $f_i^*$ is obtained using a collision operator $\Omega_i$ and an optional source term $S_i$.
Furthermore, the propagation step with $\Delta t = \Delta x = 1$ is given by
\begin{equation}
    \label{eq:stream}
    f_i(\boldsymbol{x} + \boldsymbol{c}_i \Delta t, t+\Delta t) = f_i^*(\boldsymbol{x}, t),
\end{equation}
which streams the particle populations to their corresponding neighboring lattice nodes.
\par
The simplest way to account for collisions is to relax the distributions toward their equilibrium $f_i^\text{eq}$, as is done by the Bhatnagar--Gross--Krook (BGK) collision operator \cite{Bhatnagar_Gross_Krook_1954}
\begin{equation}
    \label{eq:bgk}
    \Omega_i(\boldsymbol{x}, t) = - \omega (f_i(\boldsymbol{x}, t) - f_i^\text{eq}(\rho_\text{f}, \boldsymbol{u}_\text{f})),
\end{equation}
with the relaxation frequency $\omega$ that determines the speed of the relaxation and depends on the effective kinematic viscosity as follows
\begin{equation}
    \label{eq:relaxation}
    \omega = \left( 3 \frac{\eta^*}{\rho_\text{f}} + \frac{1}{2} \right)^{-1}.
\end{equation}
The Maxwell--Boltzmann distribution quantifying the equilibrium state reads~\cite{Qian1992}
\begin{equation}
    \label{eq:maxbolt}
    f_i^\text{eq}(\rho_\text{f}, \boldsymbol{u}_\text{f}) = w_i \rho_\text{f} \left( 1 + \frac{\boldsymbol{c}_i \cdot \boldsymbol{u}_\text{f}}{c_s^2} + \frac{(\boldsymbol{c}_i \cdot \boldsymbol{u}_\text{f})^2}{2c_s^4} - \frac{\boldsymbol{u}_\text{f}^2}{2c_s^2} \right).
\end{equation}
The required weights $w_i$ originate from a Gauss-Hermite quadrature rule and are fixed for the chosen velocity set, as is the constant lattice speed of sound $c_s$.
For D$3$Q$19$, the latter is $c_s=1/\sqrt{3}$.

\subsection{Integration of the truncated power law model}
\label{sec:methods:lbmpl}

A truncated power law model is often used to consider non-Newtonian fluids with LBM.
This ensures that the shear rate dependent local viscosity does not become too large or too small, causing instability or inaccuracy.
Consequently, for \cref{eq:hbviscmodel}, the following applies \cite{Gabbanelli_Drazer_Koplik_2005}
\begin{align}
\label{eq:truncatedpl}
    \eta^* = \max{}(\min{}(\tau_0 |\dot{\gamma}|^{-1} + k |\dot{\gamma}|^{n-1}, \eta_\text{max}), \eta_\text{min}).
\end{align}
The maximum and minimum viscosity $\eta_\text{max}$ and $\eta_\text{min}$ are predefined.
In the following studies, we use $\eta_\text{max} = 3$ and $\eta_\text{min} = 3\cdot10^{-3}$.
We emphasize that these are lattice units.
\par
Therefore, the only unknown in \cref{eq:truncatedpl} is the shear rate.
To calculate it, it is advantageous that for non-forced LBM models the second momentum of the non-equilibrium part of the particle population $f^{\text{neq}}_i$ and the shear rate tensor $\boldsymbol{E}$ are related by \cite{Dapelo_Trunk_Krause_Bridgeman_2019}
\begin{align}
\label{eq:sheartensorrate}
    \boldsymbol{E} = - \frac{\omega}{2 \rho_\text{f} c_s^2} \sum_i f_i^\text{neq}(\rho_\text{f}, \boldsymbol{u}_\text{f}) \boldsymbol{c}_i \otimes \boldsymbol{c}_i.
\end{align}
Using the above, the shear rate is
\begin{align}
\label{eq:shearrate}
    \dot{\gamma} = \sqrt{2 \boldsymbol{E} : \boldsymbol{E}}.
\end{align}
It should be noted that this method of calculating the shear rate has the advantage of being completely local.

\subsection{Homogenized lattice Boltzmann method}
\label{sec:methods:hlbm}

In this work, we solve the particulate flow problem using the \HLBM{}~\cite{Krause_Klemens_Henn_Trunk_Nirschl_2017,Trunk_Marquardt_Thaeter_Nirschl_Krause_2018}, which uses a continuous model parameter, that can be seen as a confined permeability, $B(\boldsymbol{x}, t)~\in~[0,1]$ to map rigid particles onto the entire computational domain~\cite{Haussmann_Hafen_Raichle_Trunk_Nirschl_Krause_2020,Trunk_Weckerle_Hafen_Thaeter_Nirschl_Krause_2021,hafen2023NumericalInvestigationDetachment}.
By using the signed distance to the surface of the particle in this level set function, the method accommodates particles of any given shape~\cite{Marquardt2024a}.
\par
The exact difference method (EDM) introduced by Kupershtokh et al. \cite{kupershtokh2009EquationsStateLattice} is adapted to account for the influence of the particles on the fluid ~\cite{Trunk_Weckerle_Hafen_Thaeter_Nirschl_Krause_2021}, which introduces the following source term in \cref{eq:collide}
\begin{equation}
	\label{eq:kupershtokh}
	S_i(\boldsymbol{x},t) = f_i^\text{eq}(\rho_\text{f}, \boldsymbol{u}_\text{f}+\Delta\boldsymbol{u}_\text{f}) - f_i^\text{eq}(\rho_\text{f}, \boldsymbol{u}_\text{f}).
\end{equation}
Using a convex combination of fluid and particle velocities~\cite{Trunk_Weckerle_Hafen_Thaeter_Nirschl_Krause_2021}
\begin{equation}
	\label{eq:hlbm:du}
	\Delta \boldsymbol{u}_\text{f}(\boldsymbol{x}, t) = B(\boldsymbol{x}, t) \left( \boldsymbol{u}_\text{p}(\boldsymbol{x}, t) - \boldsymbol{u}_\text{f}(\boldsymbol{x}, t) \right),
\end{equation}
the velocity difference $\Delta \boldsymbol{u}_\text{f}(\boldsymbol{x}, t)$ is obtained.
The coupling from the fluid to the particles follows Wen et al. \cite{wen2014GalileanInvariantFluid}, leading to the local hydrodynamic force
\begin{align}
	\label{eq:hlbm:mea}
	\boldsymbol{F}_{\text{h}}(\boldsymbol{x}, t) = 
	\sum_i (\boldsymbol{c}_i - \boldsymbol{u}_\text{p}(\boldsymbol{x}, t)) f_i(\boldsymbol{x} + \boldsymbol{c}_i, t) + (\boldsymbol{c}_i + \boldsymbol{u}_\text{p}(\boldsymbol{x}, t)) f_{\bar{i}}(\boldsymbol{x}, t).
\end{align}
The integral hydrodynamic force is calculated by aggregating the local forces occurring at all nodes within the particle.
These nodes are represented by the position vector $\boldsymbol{x}_\text{b}$.
Thus, the resulting total hydrodynamic force is expressed as
\begin{align}
	\label{eq:hlbm:hydrodynamicforce}
	\boldsymbol{F}_{\text{p}}(t) = 
	\sum_{\boldsymbol{x}_\text{b}} \boldsymbol{F}_{\text{h}}(\boldsymbol{x}_\text{b}, t).
\end{align}
The torque is governed by
\begin{align}
	\label{eq:hlbm:hydrodynamictorque}
	\boldsymbol{T}_{\text{p}}(t) = 
	\sum_{\boldsymbol{x}_\text{b}} (\boldsymbol{x}_\text{b} - \boldsymbol{X}_\text{p}) \times  \boldsymbol{F}_{\text{h}}(\boldsymbol{x}_\text{b}, t).
\end{align}

\subsection{Discrete contacts}
\label{sec:methods:contact}

The particle-fluid coupling process performs a preliminary contact detection by evaluating the particle distances at each node.
If the distance between two particles is less than half the cell diagonal ($d_\text{s}<\sqrt{0.75}\Delta x$), a contact is assumed and an initial approximate cuboid or bounding box is formed around the contact area.
This bounding box serves as the domain for iterative calculations to determine the parameters necessary to compute the contact forces, including the overlap volume, the indentation depth, and the contact normal.
However, the relatively large lattice spacing often results in an inaccurate bounding box due to insufficient overlap resolution.
To compensate for this, a refinement step is performed by iterating over the surface of the initial approximation using a specified number of points in each spatial direction, known as the contact resolution $N_\text{c}$.
In this paper we use $N_\text{c}=7$.
The distance to the actual contact is calculated at each point in discrete directions, increasing the precision of the bounding box.
Finally, the contact resolution $N_\text{c}$ is used to iterate over the entire overlap region, determining the overlap volume, contact point, contact normal, indentation depth, and contact force.
For a complete understanding of the algorithm, please refer to the corresponding publication~\cite{Marquardt_Römer_Nirschl_Krause_2023,Marquardt2024b}.

\section{Setups}
\label{sec:setups}

\subsection{Dimensionless Numbers}
\label{sec:setups:dimensionless-numbers}
\begin{figure}[H]
	\centering
	\includegraphics{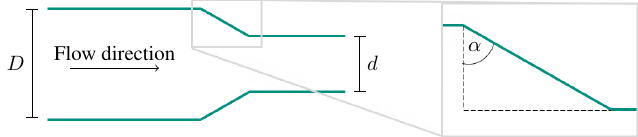}
	\caption{Sketch of the abstract geometry showing flow direction and geometric parameters.
		\label{fig:sim-geometry}}
\end{figure}

\noindent
In the context of a suspension containing fruit particles and a Herschel--Bulkley fluid flowing through a constricted cross-section as shown in~\cref{fig:sim-geometry}, the following dimensionless quantities are derived using the Pi theorem:
\begin{multicols}{2}
\begin{itemize}
    \item $\Pi_1 = \text{Re} = \frac{\rho_\text{f} D^n \bar{u}^{2-n}}{K} $
    \item $\Pi_2 = \alpha$
    \item $\Pi_3 = n$
    \item $\Pi_4 = \nu$
    \item $\Pi_5 = \phi_\text{p}$
    \item $\Pi_6 = \frac{\Delta p}{\bar{u}^2 \rho_\text{f}}$
    \item $\Pi_7 = \frac{D_\text{p}}{D}$
    \item $\Pi_8 = \frac{d}{D} = 0.5$
    \item $\Pi_9 = \frac{\rho_\text{p}}{\rho_\text{f}} = 1$
    \item $\Pi_\sigma = \frac{\sigma}{\bar{u}^2 \rho_\text{f}}$
    \item $\Pi_F = \frac{F}{\bar{u}^2 D^2 \rho_\text{f}}$
\end{itemize}
\end{multicols}
\noindent
$\Pi_1$ is the Metzner--Reed Reynolds number~\cite{metzner1955FlowNonnewtonianFluids}.
The last two mentioned dimensionless quantities apply to all stresses $\sigma$ and elasticity modules $E$ ($\Pi_\sigma$) as well as forces $F$ ($\Pi_F$).
To avoid repetition, however, the quantities have only been listed once.
Furthermore, $\bar{u}$ refers to the average velocity of the suspension, $\alpha$ to the taper angle, and $D_\text{p}$ to the particle size as the diameter of a volume-equivalent sphere.

\subsection{Experimental}
\label{sec:setups:experiment}
\paragraph{Preparation of the model fluids}
The preparation of the model fluids is done in three steps. The preparation of liquid and the particulate phase, followed by a mixing step. The liquid phase consists of $23\%$ sucrose, $1.169\%$ NaCl, $0. 924\%$ sodium carboxymethyl cellulose, and $0.0616\%$ Laponite  as a rheology additive to implement a yield point. First, the sucrose, the carboxymethyl cellulose, and the NaCl are homogenized into distilled water at $60~\si{\celsius}$. The carboxymethyl cellulose solution is then left  $24~\si{\hour}$  for hydration. Then, the Laponite solution is blended in a ratio of $3 : 2$.
\\
The particulate phase is represented by hydrocolloid gels from $1.106\%$ $\kappa$-Carrageenan and $1.169\%$ NaCl to imitate the material characteristics of the fruit pieces. 
$20.5\%$ Sucrose is added to adjust the density and refractive index matching, necessary for the usage in PIV experiments. 
To prepare the gels, the mixture was heated for $30~\si{\minute}$ to $65~\si{\celsius}$ and subsequently
stored for gelation in $5~\si{\liter}$ buckets for $24~\si{\hour}$ at $7~\si{\celsius}$. 
After the gelation the gel is cut into cubes of approximately $10 \times 10 \times 10 ~\si{\milli\meter}$ and carefully mixed by hand with the liquid phase, adjusting the particle mass fraction to $30\%$.

\paragraph{Rheological Analysis of the liquid phase}
Firstly, the yield points of the liquid phases are determined through an amplitude sweep. For this purpose, an MCR 102 Rheometer (Anton Paar GmbH, Graz, Austria) with a CP50-1 cone-plate measuring geometry is used. The measurement is conducted under shear rate control, with a shear strain ranging from $0.01\%$ to $100\%$ at a frequency of $f = 10~\si{\radian\per\second}$. After determining the yield points, flow curves of the samples are created. Hence, the samples are measured utilizing a controlled shear rate protocol consisting of four sections. In the first section, pre-shearing is conducted for $60~\si{\second}$ at a shear rate of $\dot{\gamma}=0.1~\si{\per\second}$. In the second section, a linear shear rate ramp from $0.1$ to $250~\si{\per\second}$ is applied. The third section involved a holding time of $60~\si{\second}$ at a shear rate of $250~\si{\per\second}$, while in the fourth section, a downward ramp from $250$ to $0.1~\si{\per\second}$ is applied. The evaluation of the measurement data is performed in MATLAB (The MathWorks, Inc, Natick, USA). Here, the measurement data of the downward ramp is utilized to determine the parameters of the nonlinear Herschel--Bulkley model, see \cref{eq:hbviscmodel}, using the method of least squares.

\paragraph{Mechanical Testing}
For optical flow measurements, imitates of the fruit pieces were created. These imitates showed a similar mechanical behavior as peach samples from fruit preparations without mimicking the anisotropic behavior \cite{eysel2021,eysel2023}.
These hydrogels allow an extensive mechanical testing, since defined probes can be created. This is used to determine the end of the linear elastic range and the end of the viscoelastic range in uniaxial compression that is needed for the determination of the damage potential as described in \ref{sec:setups:simulation}.
To characterize the mechanical behavior of the hydrogel samples, an
uniaxial compression test is performed with a Z 1.0 (Zwick/Roell GmbH \& Co. KG, Ulm, Germany) testing machine. 
A 50 mm diameter compression plate is attached to a Xforce P (5 N) force transducer with a maximum load of $1~\si{\kilo\newton}$.
The tests are carried out with a preload force of $0.1~\si{\newton}$. A sheet of paper is placed under the sample, creating friction between the material and the rigid support to avoid slipping.
Each sample is compressed to failure to register its strength and strain.
A preload force of $0.1~\si{\newton}$ is reached with a velocity
of $15~\si{\milli\meter\per\minute}$. Then, the gel sample is compressed with a speed of $5~\si{\milli\meter\per\minute}$ until the desired tool distance is reached.

\begin{figure}[H]
    \centering
    \includegraphics[width=0.7\textwidth]{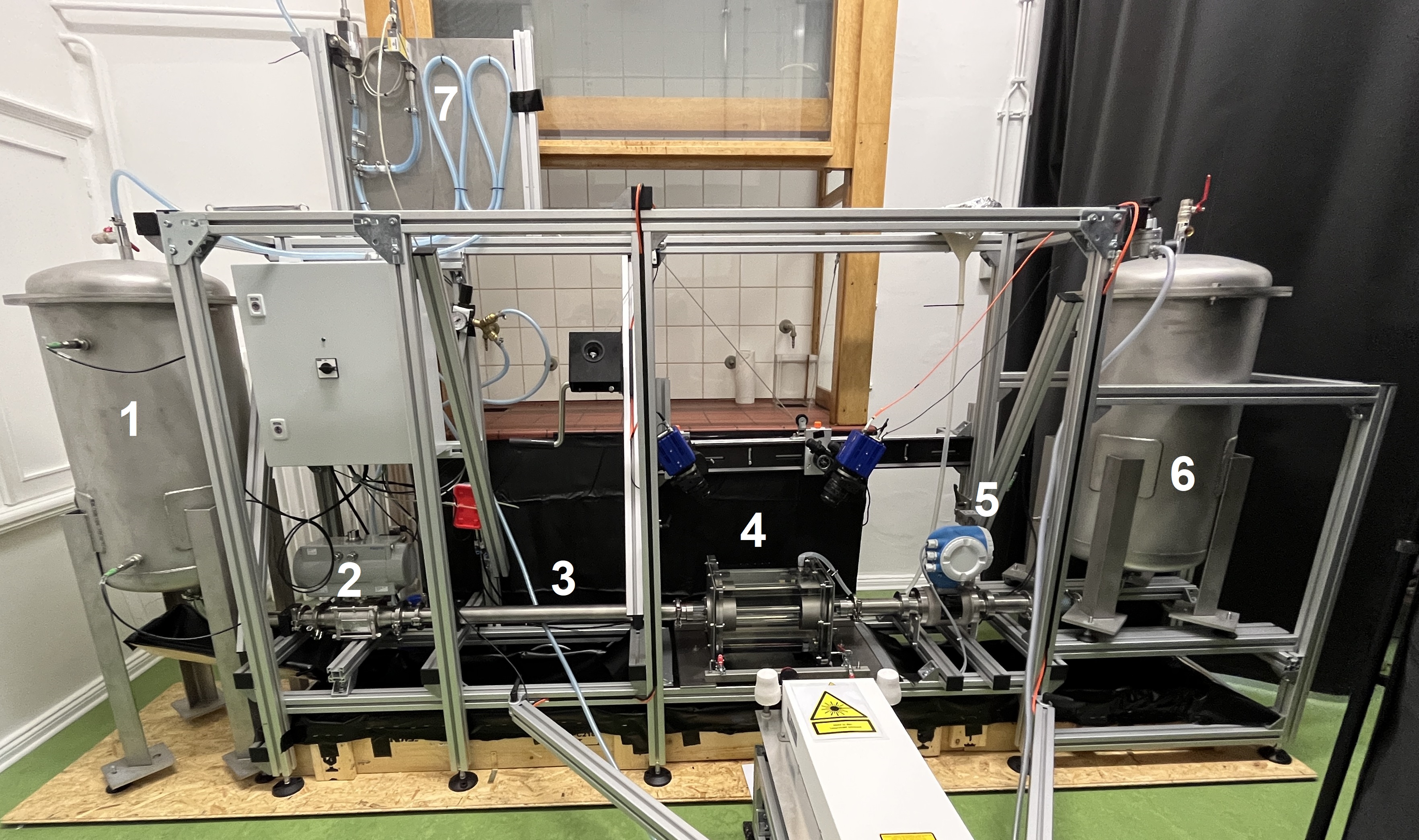}
    \caption{Experiment for the evaluation of the damage potential of different pipe geometries.\label{fig:experimental-setup}}
\end{figure}

\paragraph{Flow channel} The experimental setup shown in \cref{fig:experimental-setup} consists of tank 1 (1), a pneumatic ball valve (2) (FESTO, Esslingen am Neckar, Germany), an inlet section (3) with a length of $75.11~\si{\centi\meter}$, the measurement geometry (4), a magnetic-inductive flow meter (5) (MID) of the type Promag H300 (Endress + Hauser, Reinach, Switzerland), and tank 2 (6). The two tanks each have a total volume of $100~\si{\liter}$. They are equipped with two level sensors to prevent overfilling and air ingress into the pipeline. For this purpose, the level sensors are coupled to the ball valve via a relay circuit. The ball valve is only used to open and close the pipeline and not as a control valve. The flow rate is controlled by pressurizing tank 1 using a PID controller. For this purpose, an EL-PRESS pressure measuring and control valve (7) (Bronkhorst High-Tech B.V., Ruurlo, Netherlands) is integrated in a cascade circuit in LabVIEW (National Instruments, Austin, United States). The experiments are recorded with a Mini Ax 100 (Photron, Tokyo, Japan) high-speed camera operated in the DaVis (LaVision GmbH, Göttingen, Germany) PIV imaging software.

\begin{figure}[H]
    \centering
    \includegraphics[width=\textwidth]{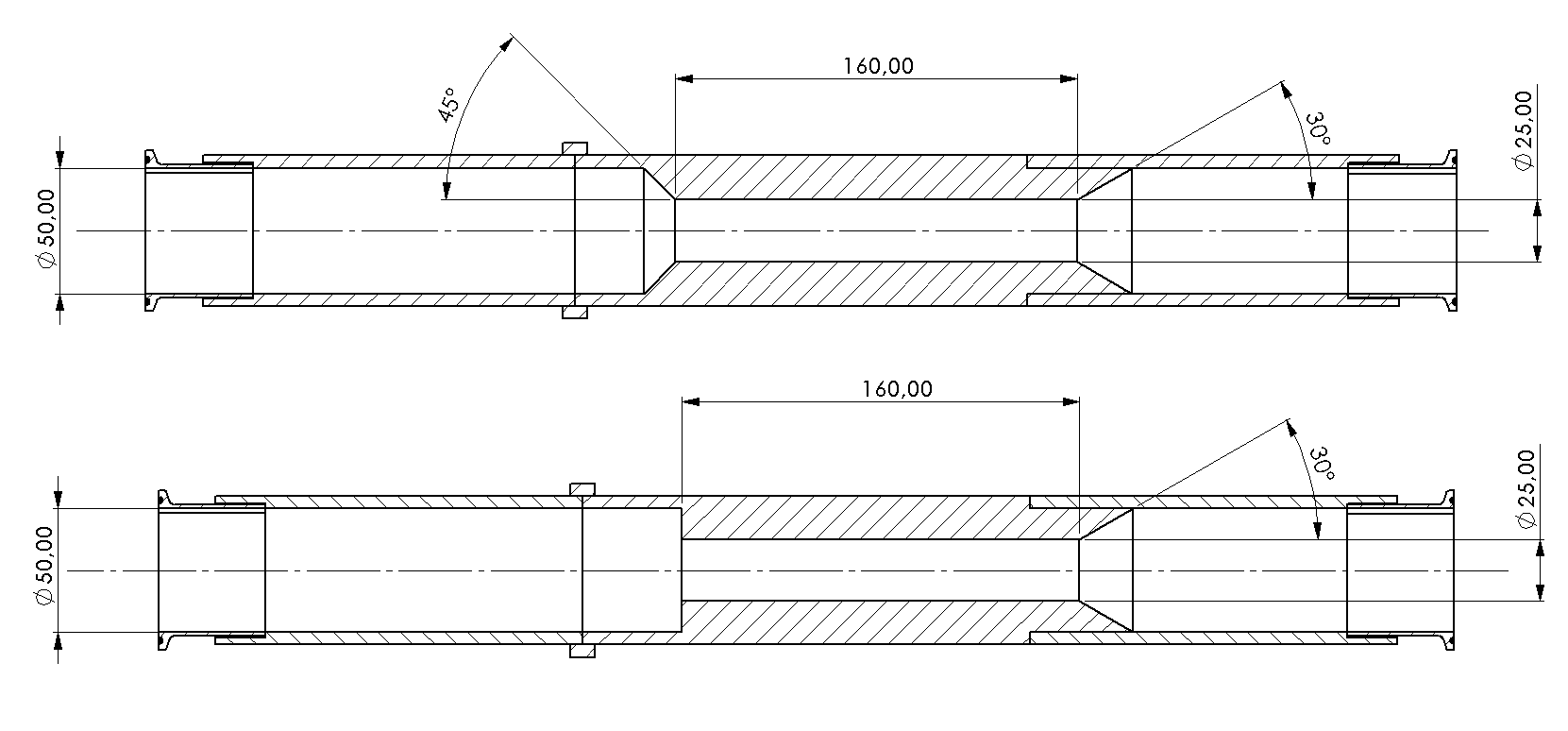}
    \caption{Experiment for the evaluation of the damage potential of different pipe geometries.\label{fig:pipe-geometries}}
\end{figure}

\paragraph{Pipe Geometries} A total of three pipe geometries are used. In order to validate the numerical simulations, a straight pipe with a diameter of $50~\si{\milli\meter}$ and a sudden and gradual cross-sectional change from $50~\si{\milli\meter}$ to $25~\si{\milli\meter}$ are investigated.
Sketches of the pipe geometries are shown in \cref{fig:pipe-geometries}.

\paragraph{Flow Experiments} Approximately $50~\si{\liter}$ of industrial fruit preparation and model fluid are passed through the geometries in 20 runs each. Particle size distribution is determined before and after the experiments in triplicates. For each experiment, a fresh sample is taken, and the fruit preparations are conveyed at rates of $20~\si{\liter\per\minute}$ and $40~\si{\liter\per\minute}$. Thus, Reynolds numbers of $Re=2$ and $Re=5$ are set for pipe diameters of $50~\si{\milli\meter}$. Due to high effort for their preparation, the model fluids are conveyed at volume flow rates of $20~\si{\liter\per\minute}$, to achieve $Re=5$. 

\begin{figure}[H]
\centering
\begin{subfigure}{0.45\textwidth}
\includegraphics[width=\linewidth]{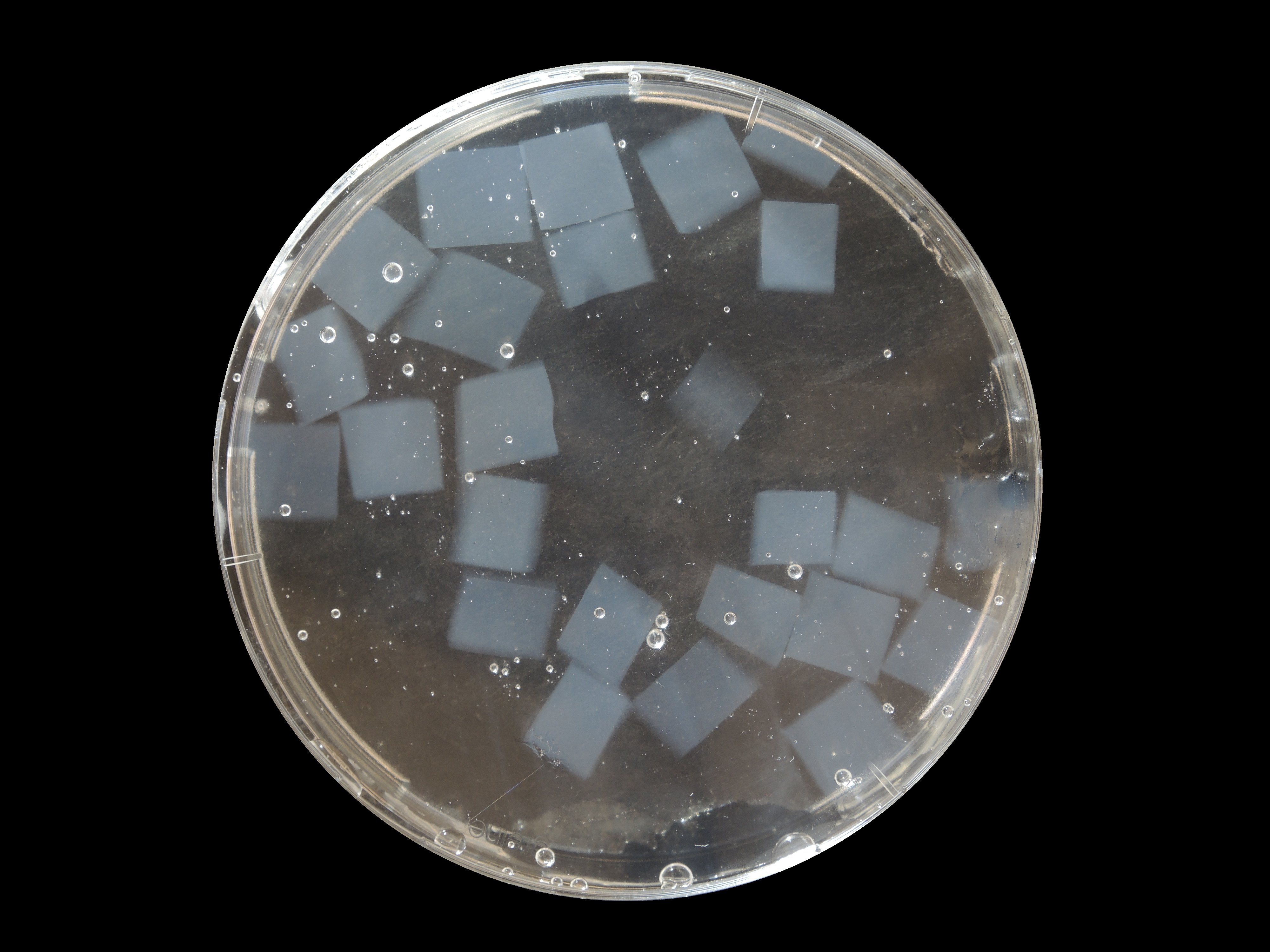}
\caption{Original image}
\label{fig:ImJ_original}
\end{subfigure}
\begin{subfigure}{0.45\textwidth}
\includegraphics[width=\linewidth]{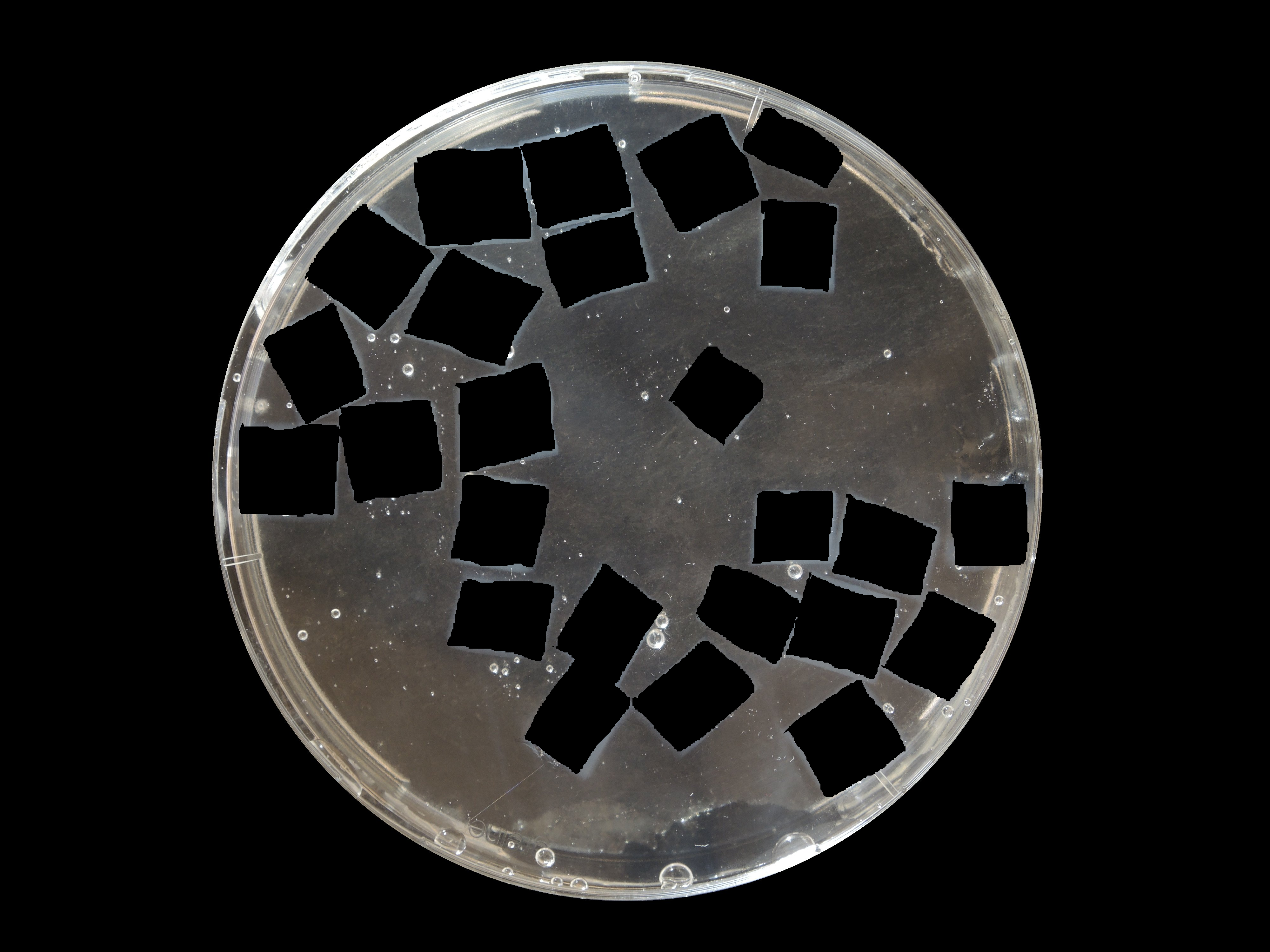}
\caption{Cropped image}
\label{fig:ImJ_Crop}
\end{subfigure}

\caption{Example for the measurement of the model particles before and after the experiments.
\label{fig:images_ImJ}}
\end{figure}

\paragraph{Determination of particle size distribution} The particle size distribution of the fruit preparations is determined in triple determination using wet sieving. Sieving is performed using an Analysette sieve tower (Fritsch GmbH, Idar-Oberstein, Germany) and six sieves (mesh sizes:~$8,~6.3,~5,~4,~2,$~\text{and}~$1~\si{\milli\meter}$). $500~\si{\gram}$ of the fruit preparation is sieved with an amplitude of $2~\si{\milli\meter}$ for three minutes. The material is sieved wet for two minutes and dry for one minute. Spraying, at a flow rate of $1.4~\si{\liter\per\minute}$ at $40~\si{\celsius}$, is achieved using a sieve cover with three spray nozzles. The liquid is drained through a sieve tray with an outlet. Subsequently, the sieves are set up at an angle of $\ang{45}$ for five minutes to allow the remaining water to drain off. The weight of each fraction is then determined.\\
The particle size distribution of the model particles is determined by the analysis of images of samples in Petri dishes. For every conveying experiment, 20 Petri dishes are filled with sample and pictures are recorded using a Nikon COOLPIX P7700 (Nikon, Tokio, Japan) digital camera. The upper surfaces of the containing particles are determined using ImageJ (NIH, Bethesda, USA) where the measured area is cropped for traceability reasons. Altogether $1741$ particles are measured, an example of the procedure is shown in \cref{fig:images_ImJ}. The number of measured particles ($n_\text{p}$) ranges from $417-467$ per experiment.

\subsection{Computational}
\label{sec:setups:simulation}

In this section, the simulation setup is described, focusing on the geometry, studied parameters, discretization, boundary and initial conditions and  employed.

\paragraph{Geometry}
The simulations focus on the constriction of the cross-section.
The diameter at the inlet is set to $D=50~\si{\milli\meter}$ and is reduced to $D/2$ in the second part.
The transition section is characterized by the taper angle $\alpha$, as shown in \cref{fig:sim-geometry}.
Here, $\alpha = \ang{0}$ represents an abrupt change in cross-section, while $\alpha=\ang{90}$ describes a straight pipe.
The total length of the pipe is $10D$, with the left part comprising 70\% to ensure sufficient particle delivery.
For $\alpha<\ang{90}$, the conical narrowing length is given by $D/4 \tan{\alpha}$.

\paragraph{Physical Parameters}
The main parameters of investigation are the Reynolds number $\Rey\in\{2, 3, 5, 8\}$ and the taper angle $\alpha~\in~\{0, \frac{\pi}{8}, \frac{\pi}{4}, \frac{3\pi}{8}, \frac{\pi}{2}\}$.
The decision to focus on these specific parameters is based on the practical importance of avoiding changes in the formulation of the product for consumer acceptance.
The mechanical properties of fruit preparations depend on a number of factors, including season, degree of processing, and recipe.
As a result, they are not easily adjusted.
Varying the Reynolds number, on the other hand, allows analysis of the effects of different speeds on stress, making it a crucial adjustable parameter during plant design.
Similarly, the angle of cross-sectional change can be adjusted during design.
The non-Newtonian fluid is modeled as a Herschel--Bulkley fluid, with flow index $n = 0.42$, consistency $K = 13.1~\si{\pascal\second\tothe{0.42}}$, and density $\rho_\text{f} = 1100~\si{\kilogram\per\cubic\meter}$.
Particles are immersed in the fluid with a particle volume fraction $\phi_\text{p}~\approx~0.3$.
The cubic particles vary in size, i.e. $\Pi_7~\in~\{0.159,0.2\}$, and have a Poisson's ratio $\nu = 0.3$, consistent with data reported on apples~\cite{Finney1967}, and two Young's moduli of $30~\si{\kilo\pascal}$ and $60~\si{\kilo\pascal}$, corresponding to those of the model particles and the considered fruit particles.
The steel walls have an elastic modulus of $190~\si{\giga\pascal}$ and a Poisson's ratio of $0.3$~\cite{Chen2016}.
Gravity is neglected due to the similar density of the fluid and particles, and $\rho_\text{p} = \rho_\text{f}$ is assumed.
The restitution coefficient in the simulations is $0.9$, the static friction coefficient is $0.15$~\cite{Yildiz2014}, and the sliding friction coefficient is $0.1$.

\paragraph{Discretization}
During the simulation, the inlet diameter is resolved with $N$ grid points, thus $\Delta x = D/N$.
In case of contacts, the contact region resolution is $N_\text{c} = 7$.
The time step size is $\Delta t = 5 ~\si{\micro\second}$.

\paragraph{Boundary Conditions}
A constant velocity field is prescribed at the inlet, with no-slip conditions at the walls and a constant pressure at the outlet.
To obtain the velocity profile, a preliminary simulation in a straight pipe of length $6D$ is conducted, with a constant inlet velocity corresponding to the mean flow velocity $\bar{u}$.
The established flow profile at $x = 4.5D$ is used as a boundary condition for subsequent simulations with the same fluid parameters.
The velocity and pressure boundaries are realized using the regularized method according to Latt and Chopard~\cite{lattStraightVelocityBoundaries2008}, while the no-slip boundary condition on the curved stationary wall follows Bouzidi et al.~\cite{bouzidi2001MomentumTransferBoltzmannlattice}.

\begin{figure}[H]
	\centering
	\includegraphics[width=\textwidth]{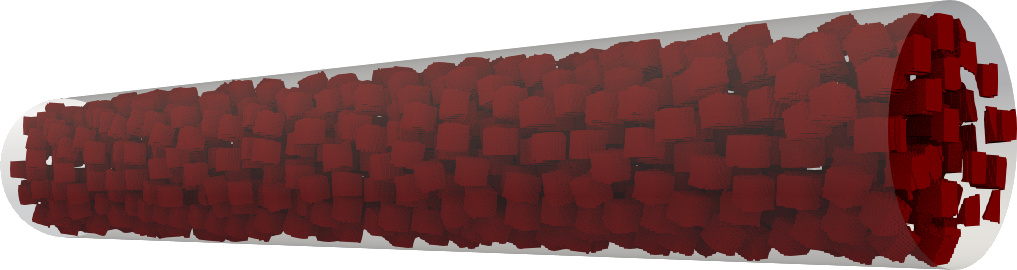}
 \ \\
 \ \\
    \includegraphics[width=\textwidth]{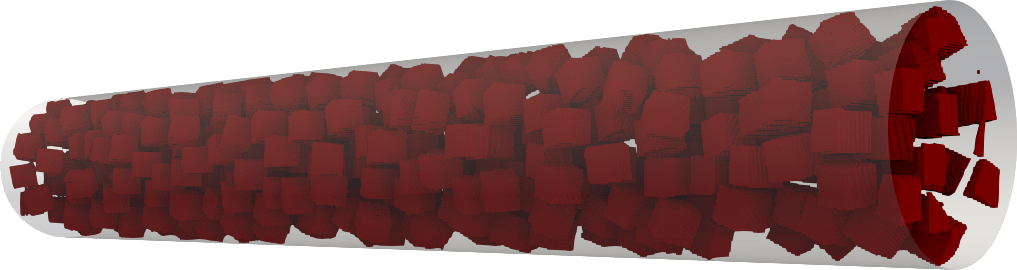}
	\caption{Visualization of the initial fruit particle distributions using a resolution $N=131$ for $\Pi_7=0.159$ (top) and $\Pi_7=0.2$ (bottom).
		\label{fig:sim-initial-particle-distribution}}
\end{figure}

\paragraph{Initial Conditions}
To limit the computational intensity of particle-laden simulations, the fluid velocity field is first computed without particles and then used as initial conditions for simulations including particles.
Particles inherit the fluid velocity at their center of mass.
The total simulation time is $1.1~\si{\second}$, with the first $0.1~\si{\second}$ used to smooth initial gradients resulting from the particles being initialized without rotation.
Since the initial particle distribution is unpredictable, random particle positions are generated to achieve a homogeneous distribution, as observed in experiments from tanks and desired in industrial applications.
The initial distributions in a periodic pipe of length $7D - D_\text{p}$ are shown in \cref{fig:sim-initial-particle-distribution}.
The length is chosen to match the front section's length, with the reduction by the particle diameter avoiding significant overlaps with boundaries at the start of the subsequent simulations.

\paragraph{Damage potential}
In the following, the potential for damage to the particle component is quantified by
\begin{equation}
    DP = \frac{1}{n} \sum_{j=1}^{n} 	
    \begin{cases}
        1 &\text{for } \sigma_j \geq \sigma_\text{v} \\
		  \frac{\sigma_\text{v} - \sigma_j}{\sigma_\text{v} - \sigma_\text{l}} &\text{for } \sigma_\text{l} \leq \sigma_j < \sigma_\text{v} \\
		0 &\text{for } \sigma_j < \sigma_\text{l}
	\end{cases}.
\end{equation}
The damage potential indicates the probability of damage.
For this purpose, the probability of damage per data point $j$ is determined.
Subsequently, these are summed up and divided by the total number of data points $n$.
Here, the experimental data described in Section~\ref{sec:setups:experiment} and shown in \cref{tab:hydrogel} is used.
For stresses within the linear elastic range ($\sigma_j < \sigma_\text{l}$), it is assumed that no damage occurs, above the viscoelastic range ($\sigma_j \geq \sigma_\text{v}$), damage is always assumed to occur.
In the viscoelastic range, the gradual transition is approximated by a linear progression.
The aforementioned damage potential is evaluated for each type of stress, i.e. normal and tangential stress due to interaction with the fluid as well as normal and tangential stress due to contact with other particles and walls.

\begin{figure}[H]
	\centering
	\includegraphics[height=6.5cm,width=0.95\textwidth]{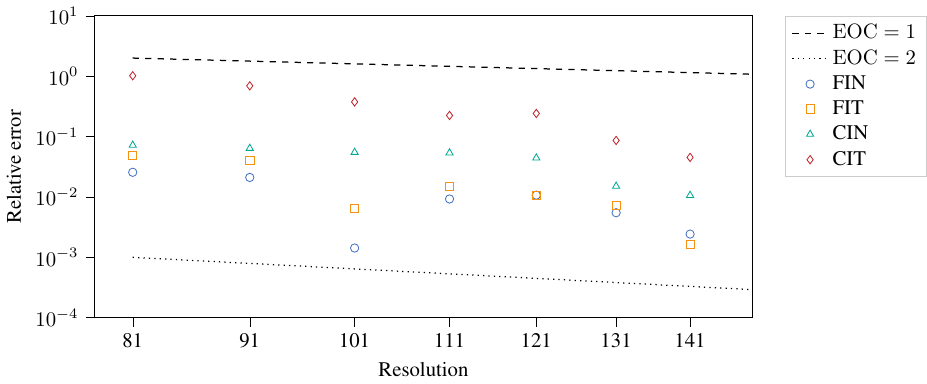}
	\caption{Relative error of the FIN, FIT, CIN, and CIT damage potential in $L^2$ norm versus the reference resolution $N=151$.
		\label{fig:grid_independence}}
\end{figure}

\paragraph{Grid independence}
In the grid independence study we consider $\alpha~=~\ang{45}$, $\Rey~=~5$, $\Pi_7~=~0.2$, $E~=~30~\si{\kilo\pascal}$, and $N \in \{81,91,101,111,121,131,141\}$.
These resolutions are compared to a baseline resolution of $N=151$ to confirm grid independence.
The results are visualized in \cref{fig:grid_independence}, which shows the relative error of the damage potentials, calculated using the $L^2$ norm as described in \cite{Krueger2016}, plotted against the different grid resolutions.
In the figure, circles represent the fluid-induced normal (FIN) stress, squares represent the fluid-induced tangential (FIT) stress, triangles represent the contact-induced normal (CIN) stress, and diamonds represent the contact-induced tangential (CIT) stress as the corresponding causes of the damage potential.
In addition, the figure includes lines denoting experimental orders of convergence with values of $1$ and $2$.
It can be seen that, in general, the error decreases with increasing resolution, more closely following the line for $\mathrm{EOC} = 2$.
The relative error falls below 10\% for all damage potentials for $N~\geq~131$.
Therefore, $N~=~131$ is chosen for the following studies.
It is also noteworthy that the error for the damage potential due to CIT stresses is generally higher due to their lower values, as can be seen in Section~\ref{sec:results:comp}.

\section{Results}
\label{sec:results}

This section presents the results of our research, including both experimental and computational results.
We begin in Section~\ref{sec:results:exp} with the results of our experiments, providing an analysis of the data collected.
Then, in Section~\ref{sec:results:comp}, we turn our attention to the computational results, exploring the results of our simulations and their implications.
Next, in Section~\ref{sec:results:validation}, we validate our experimental and computational results by comparison, establishing the reliability and consistency of our findings.
In the discussion that follows in Section~\ref{sec:results:discussion}, we interpret the results, compare experimental and computational data, and consider the broader context of our findings within the field.

\subsection{Experimental}
\label{sec:results:exp}

\paragraph{Rheological Analysis of the liquid phase}
 The fitting of the Herschel--Bulkley model resulted in the following parameters for the fluid of the Fruit preparations. The yield stress is $\tau_0 = 0.653~\si{\pascal}$, the flow index is $n = 0.42$, and the consistency is $K = 13.1~\si{\pascal\second\tothe{0.42}}$. 
 The yield stress of the model fluid is $\tau_0 = 0.244~\si{\pascal}$, the flow index is $n = 0.512$, and the consistency is $K = 2.934~\si{\pascal\second\tothe{0.512}}$. 
\paragraph{Mechanical Testing}

The results of the mechanical testing are shown in \cref{tab:hydrogel}. The stress limits of the linear elastic and viscoelastic ranges are used to determine the damage potential as described in Section~\ref{sec:setups:simulation}.

\begin{minipage}{\textwidth}
\centering
\begin{table}[H]
\caption{Mechanical behavior of the hydrogel.\label{tab:hydrogel}}
\centering
\begin{tabular}{llllllllll}
\hline

\textbf{Limit}         & \textbf{Strain in $\boldsymbol{\%}$} & \textbf{Stress in $\SI{}{\mathbf{\kilo\pascal}}$}  \\ 
\textbf{Linear elastic} & $6.50\pm0.47$   & $2.228\pm0.258$  \\
\textbf{Viscoelastic}   & $22.06\pm2.02 $ & $13.47\pm1.60$   \\
\textbf{Failure}        & $30.58\pm2.47$  & $20.27\pm6.09$   \\ \hline

\end{tabular}
\end{table}
\end{minipage}
\\
\paragraph{Damage Investigation}
Example images of the recorded footage are shown \cref{fig:images_F_smooth,fig:images_MF_smooth}. It was observed, that the fruit preparation's turbidity increased during the experiment, hampering the identification of the particle shape. It appears that the share of small particles expanded, but it can be seen that the recordings can exclusively be used for a subjective interpretation of the particle damaging. For the objective analysis, wet sieving is necessary. As it can be seen in \cref{fig:images_MF_smooth} the distinction of the model particles is not possible from the image material. Since the model particles are designed for PIV experiments, their transparent nature obstructs the inline assessment of the particle size. Further experiments showed that wet sieving also was not applicable to the model fluids, since the hydrogel particles were washed out during the wet sieving process. Therefore, the particle size of the model particles was determined by images of the Petri dishes filled with samples, that were collected after the experiment as described in \cref{sec:setups:experiment}.
\\
\begin{minipage}{\textwidth}

\begin{figure}[H]
\centering
\begin{subfigure}{0.45\textwidth}
\includegraphics[width=\linewidth]{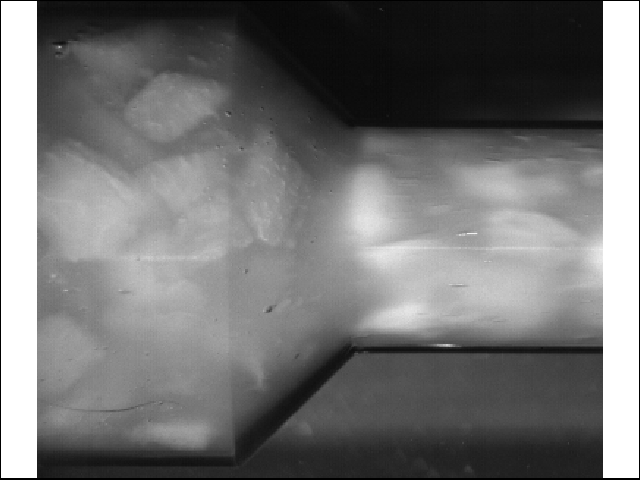 } 
\caption{Beginning of the experiment}
\label{fig:images_smooth_FZB_start}
\end{subfigure}
\begin{subfigure}{0.45\textwidth}
\includegraphics[width=\linewidth]{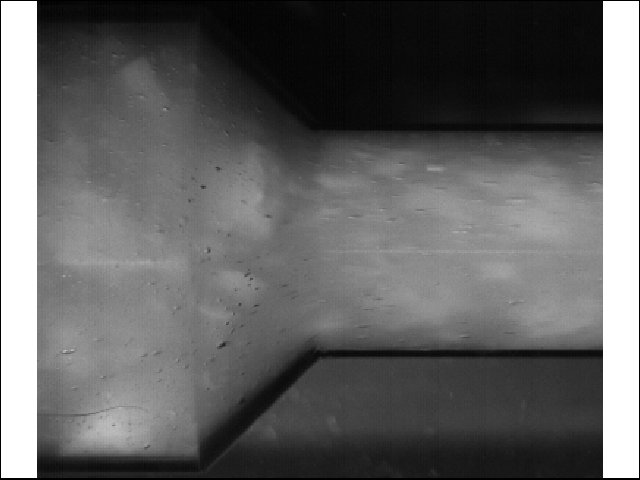}
\caption{End of the experiment}
\label{fig:images_smooth_FZB_end}
\end{subfigure}
\caption{Images taken during the conveying experiment with fruit preparations in a smooth constriction and a volume flow of $40 ~\si{\liter\per\minute}$.}
\label{fig:images_F_smooth}
\end{figure}

\begin{figure}[H]
\centering
\begin{subfigure}{0.45\textwidth}
\includegraphics[width=\linewidth]{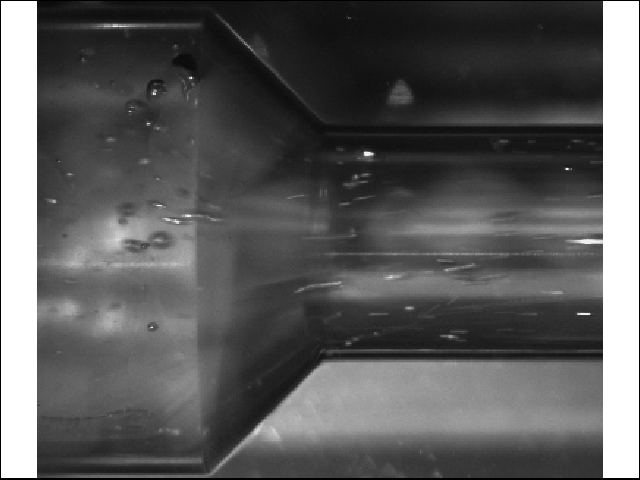}
\caption{Beginning of the experiment}
\label{fig:images_smooth_MF_start}
\end{subfigure}
\begin{subfigure}{0.45\textwidth}
\includegraphics[width=\linewidth]{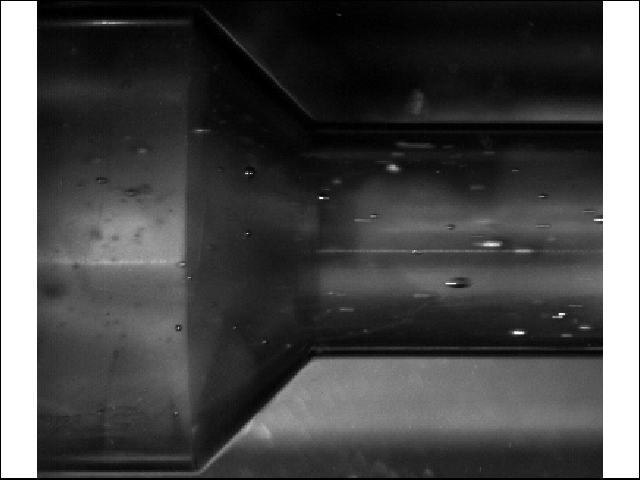}
\caption{End of the experiment}
\label{fig:images_smooth_MF_end}
\end{subfigure}

\caption{Images taken during the conveying experiment with the model fluids in a smooth constriction and a volume flow of $20 ~\si{\liter\per\minute}$.
\label{fig:images_MF_smooth}}
\end{figure}
\end{minipage}

The results of the damage investigation of the fruit preparations are shown in \cref{fig:measurement-20-before,fig:measurement-20-after,fig:measurement-40-after}. These experiments are done to compare with the numerical results for the damage potential shwon in \cref{tab:damage-potential-0.159-30,tab:damage-potential-0.159-60,tab:damage-potential-0.2-30,tab:damage-potential-0.2-60}. It can be observed that in the experiments with a flow rate of $20 ~\si{\liter\per\minute} (Re = 1.46)$, the particle size distribution is comparable to that of the unconveyed sample. The experiments with a flow rate of $40 ~\si{\liter\per\minute} (Re = 4.4)$ resulted in a significant reduction of fractions above $6.3 ~\si{\milli\meter}$, both in steady and sudden cross-sectional narrowing, with the fraction at $5 ~\si{\milli\meter}$ increasing. This cannot be seen in the straight pipe. Thus, it can be said that the change in cross-section induces damage to the fruit pieces. However, the different types of cross-sectional changes appear to have only a minor influence on the extent of damage. This confirms that the angle of narrowing has a minor impact on the damage potential of the fruit pieces. The experiments also indicate that the damage potential increases with increasing $\Pi_7=D_\text{p}/D$, leading to a decrease in fractions of larger particles.

\begin{minipage}{\textwidth}
\begin{figure}[H]
    \centering
    \includegraphics[width=0.3\textwidth]{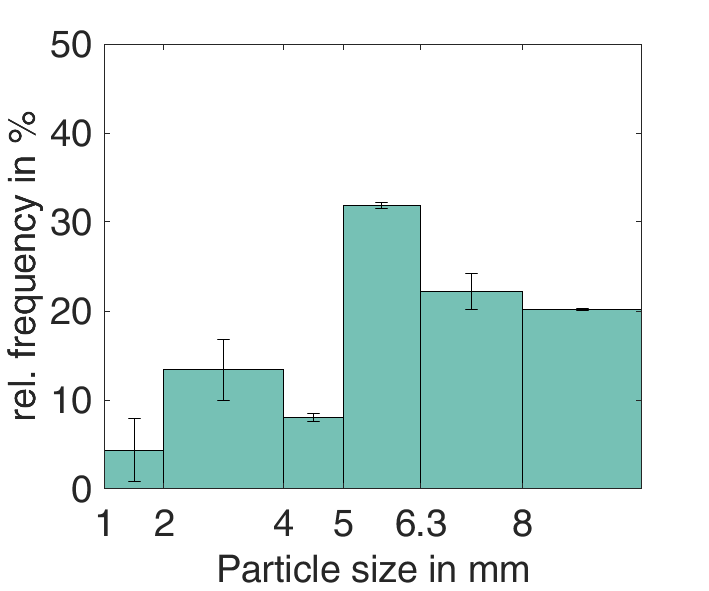}
    \caption{Initial particle distribution.\label{fig:measurement-20-before}}
\end{figure}

\begin{figure}[H]
\centering
\begin{subfigure}{0.3\textwidth}
\includegraphics[width=\linewidth]{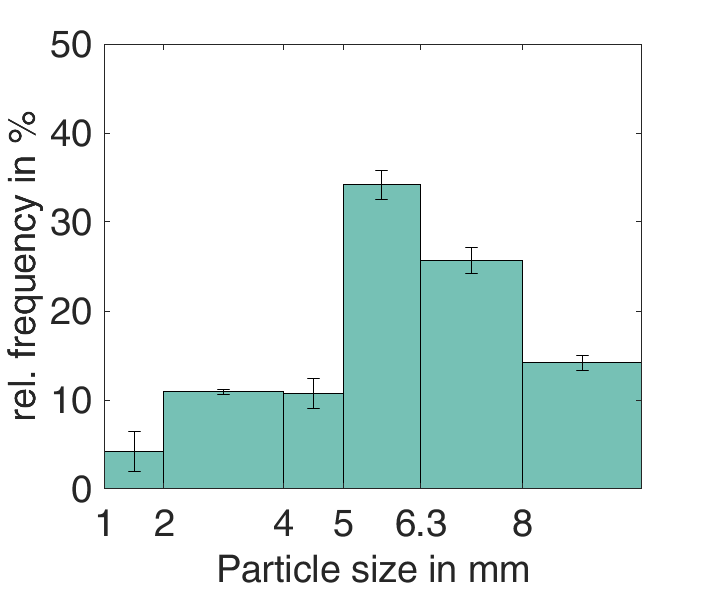} 
\caption{$\alpha=\ang{90}$}
\label{fig:measurement-20-straight}
\end{subfigure}
\begin{subfigure}{0.3\textwidth}
\includegraphics[width=\linewidth]{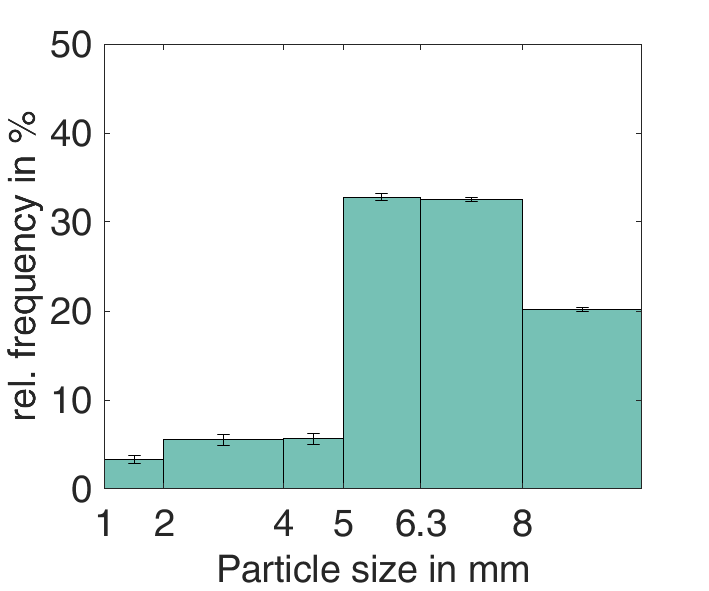}
\caption{$\alpha=\ang{45}$}
\label{fig:measurement-20-smooth}
\end{subfigure}
\begin{subfigure}{0.3\textwidth}
\includegraphics[width=\linewidth]{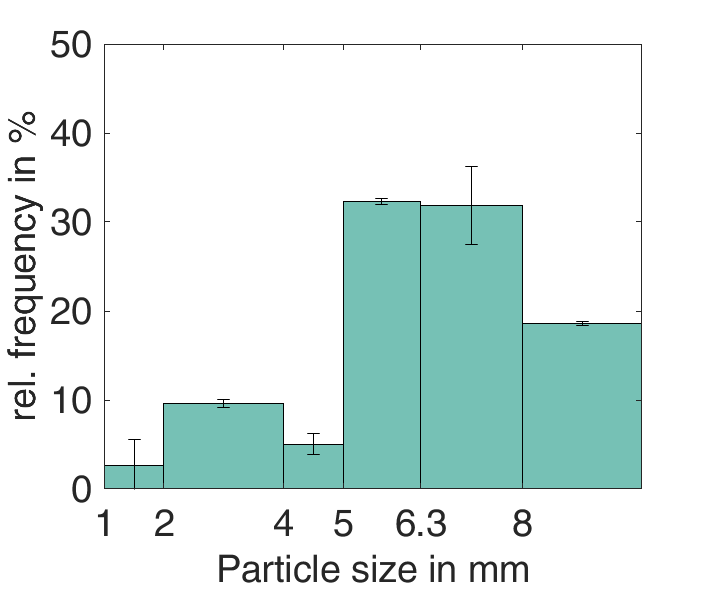}
\caption{$\alpha=\ang{0}$}
\label{fig:measurement-20-sudden}
\end{subfigure}
\caption{Particle distribution after experiments with a flow rate of $20 ~\si{\liter\per\minute}$.}
\label{fig:measurement-20-after}
\end{figure}

\begin{figure}[H]
\centering
\begin{subfigure}{0.3\textwidth}
\includegraphics[width=\linewidth]{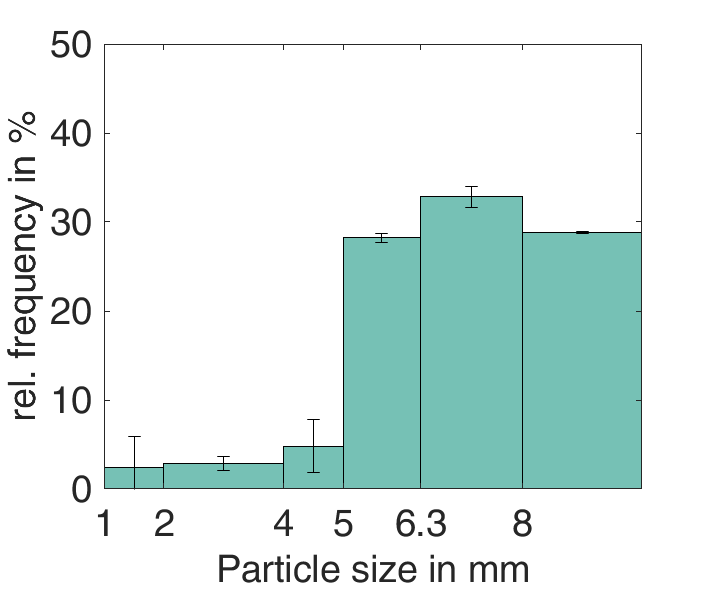} 
\caption{$\alpha=\ang{90}$}
\label{fig:measurement-40-straight}
\end{subfigure}
\begin{subfigure}{0.3\textwidth}
\includegraphics[width=\linewidth]{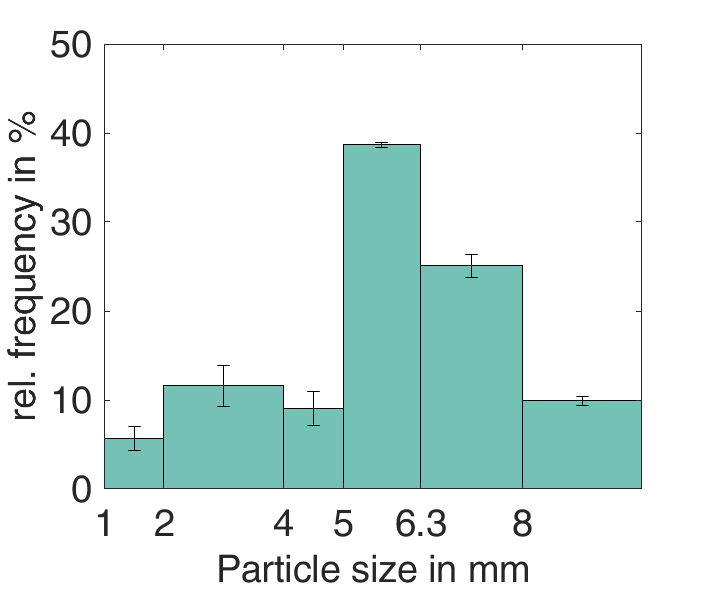}
\caption{$\alpha=\ang{45}$}
\label{fig:measurement-40-smooth}
\end{subfigure}
\begin{subfigure}{0.3\textwidth}
\includegraphics[width=\linewidth]{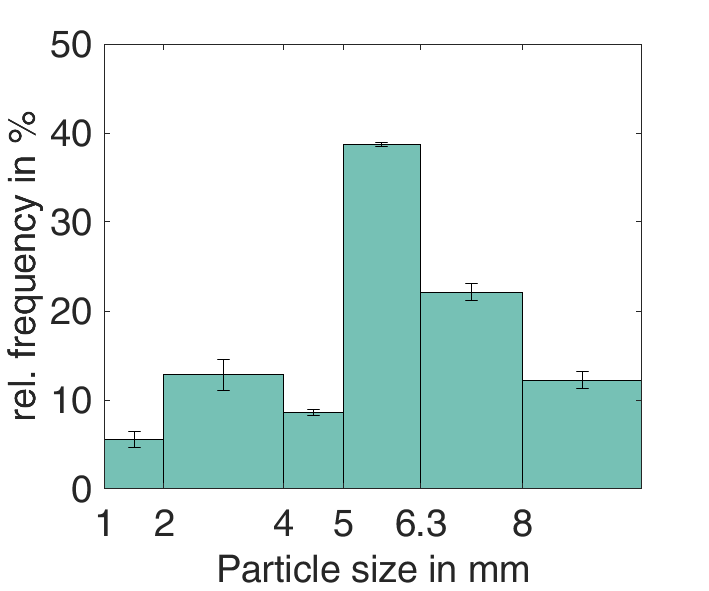}
\caption{$\alpha=\ang{0}$}
\label{fig:measurement-40-sudden}
\end{subfigure}

\caption{Particle distribution after experiments with a flow rate of $40 ~\si{\liter\per\minute}$.\label{fig:measurement-40-after}}
\end{figure}
\end{minipage}

In \cref{fig:particle-measurement-20-before,fig:particle-measurement-20-after} the results for the experiments with the model particles at a volume flow rate of $20 ~\si{\liter\per\minute} (Re = 5.78)$ are shown. Since wet sieving was not applicable on the model particles, the histograms show the area of the upper surface of the particles. The model particles show comparable damaging as the fruit particles. Since the model particles were cut by hand into cubes, they show a narrow particle size distribution, with the majority of upper surface areas between $70~-~130~\si{\square\milli\meter}$. After the conveying in the geometry with $\alpha=\ang{90}$ the fraction between $90~-~100~\si{\square\milli\meter}$ shows higher counts than before. Conveying through the geometry with $\alpha=\ang{45}$ leads to a widening of the distribution with a peak in the fraction between $90~-~100~\si{\square\milli\meter}$, whereas the geometry with $\alpha=\ang{0}$ leads to a further increased in the fraction between $90~-~100~\si{\square\milli\meter}$.

\begin{minipage}{\textwidth}
\begin{figure}[H]
    \centering
    \includegraphics[width=0.3\textwidth]{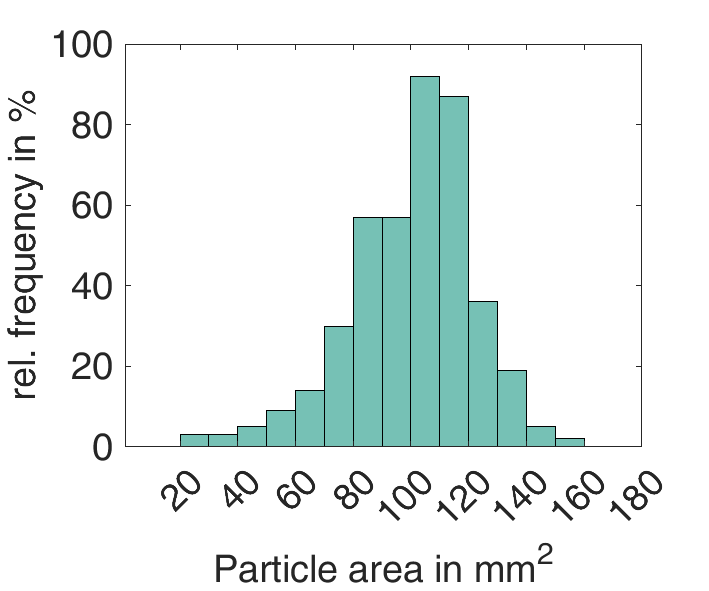}
    \caption{Initial particle distribution $n_\text{p}=419$.\label{fig:particle-measurement-20-before}}
\end{figure}

\begin{figure}[H]
\centering
\begin{subfigure}{0.3\textwidth}
\includegraphics[width=\linewidth]{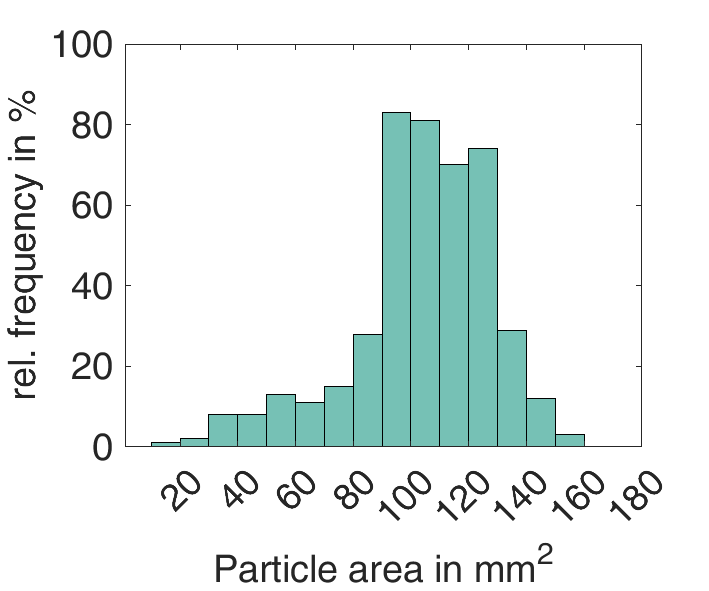} 
\caption{$\alpha=\ang{90}$, $n_\text{p}=438$}
\label{fig:particle-measurement-20-straight}
\end{subfigure}
\begin{subfigure}{0.3\textwidth}
\includegraphics[width=\linewidth]{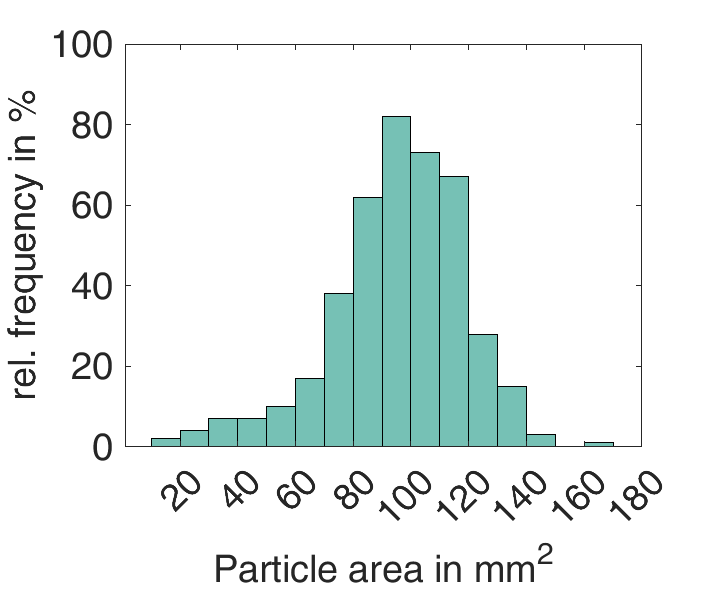}
\caption{$\alpha=\ang{45}$, $n_\text{p}=417$}
\label{fig:particle-measurement-20-smooth}
\end{subfigure}
\begin{subfigure}{0.3\textwidth}
\includegraphics[width=\linewidth]{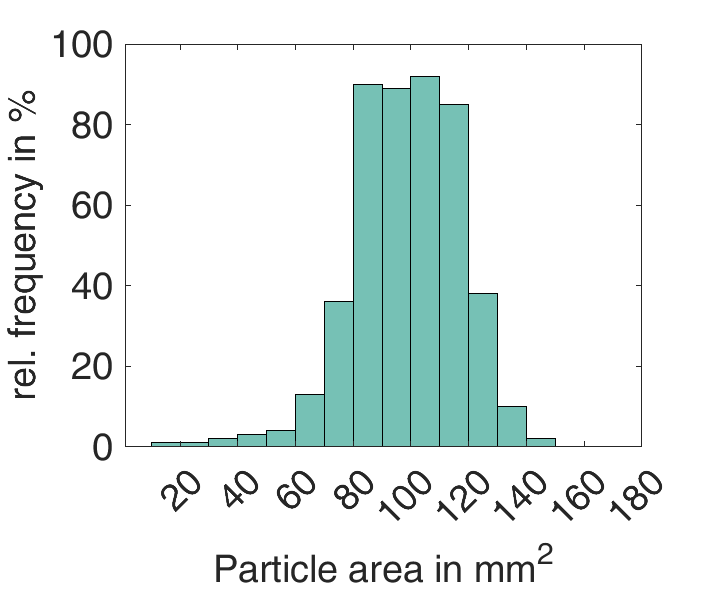}
\caption{$\alpha=\ang{0}$, $n_\text{p}=467$}
\label{fig:particle-measurement-20-sudden}
\end{subfigure}
\caption{Particle size distribution of the model particles after the experiments with a flow rate of $20~\si{\liter\per\minute}$. With $n_\text{p}$ as number of the measured particles.}
\label{fig:particle-measurement-20-after}
\end{figure}
\end{minipage}

\subsection{Computational}
\label{sec:results:comp}

\begin{figure}[H]
    \centering
    \includegraphics[height=5.45cm,width=0.6\textwidth]{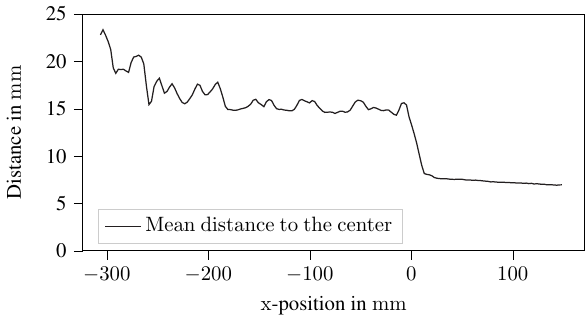}
	\caption{Plot of the average distance to the center of the pipe versus the length of the pipe.\label{fig:sim:single:plots:avg-distance}}
\end{figure}

\begin{figure}[H]
    \centering
    \includegraphics[height=5.45cm,width=0.6\textwidth]{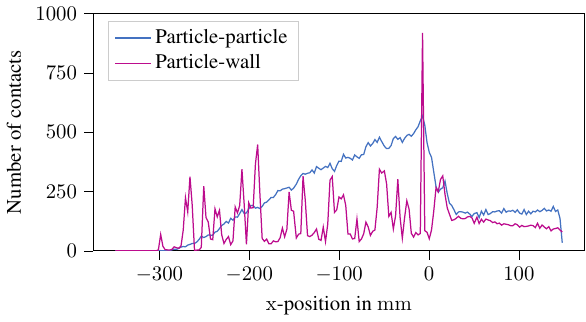}
	\caption{Plot of the number of contacts versus the length of the pipe.\label{fig:sim:single:plots:contact-frequency}}
\end{figure}

\begin{figure}[H]
    \centering
    \includegraphics[height=7.25cm,width=0.6\textwidth]{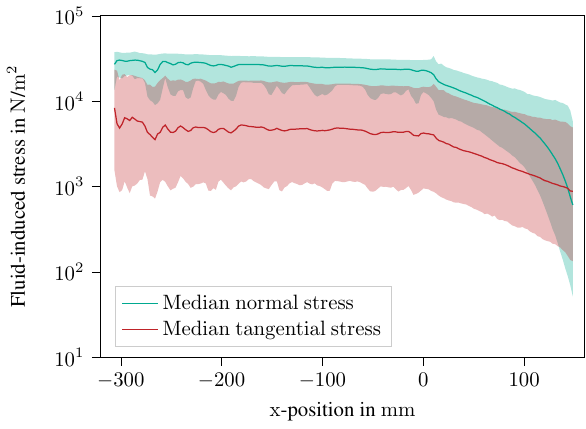}
    \caption{Plot of the fluid-induced stress  versus the length of the pipe.\label{fig:sim:single:plots:fluid-induced-stresses}}
\end{figure}

\begin{figure}[H]
    \centering
    \includegraphics[height=7.25cm,width=0.6\textwidth]{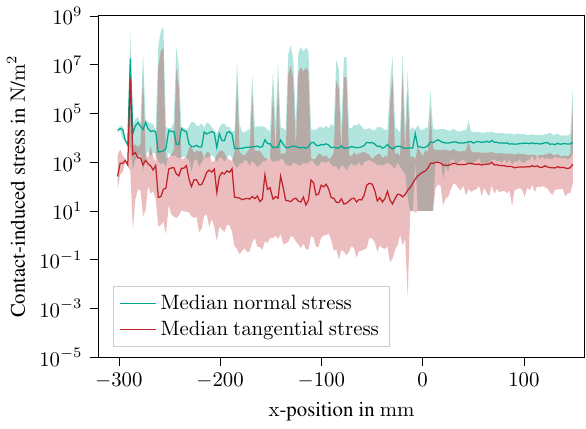}
	\caption{Plot of the contact-induced stress versus the length of the pipe.\label{fig:sim:single:plots:contact-induced-stresses}}
\end{figure}

The results of the simulation with $\Rey~=~5$, $\alpha~=~\ang{45}$, $\Pi_7~=~0.2$, $E~=~30~\si{\kilo\pascal}$, and $N =131$ are given in \cref{fig:sim:single:plots:avg-distance,fig:sim:single:plots:contact-frequency,fig:sim:single:plots:fluid-induced-stresses,fig:sim:single:plots:contact-induced-stresses}, showing the average distance to the center of the pipe in \cref{fig:sim:single:plots:avg-distance}, number of particle-particle and particle-wall contacts in \cref{fig:sim:single:plots:contact-frequency}, fluid-induced stress in \cref{fig:sim:single:plots:fluid-induced-stresses}, and contact-induced stress in \cref{fig:sim:single:plots:contact-induced-stresses} versus the length of the pipe.
The plots in \cref{fig:sim:single:plots:fluid-induced-stresses,fig:sim:single:plots:contact-induced-stresses} show the median of the induced stresses as a line next to the range that includes 90\% of the data, shown as an area.
This range excludes the 5\% lowest and 5\% highest stresses to highlight trends while minimizing the impact of outliers.

\cref{fig:sim:single:plots:avg-distance} indicates that the motion of particles is directed towards the central point of the pipe, due to the narrowing of the pipe's cross-section.
This leads to a rise in the frequency of particle-to-particle interactions, as visible in \cref{fig:sim:single:plots:contact-frequency}, as they undergo repositioning in the path leading to the constriction, and subsequently maintain a steady level of contact.
Furthermore, instances of particle-wall contacts are most prevalent at the onset of the constriction and the area preceding it, and decrease in the section with a smaller diameter.


Examining the \cref{fig:sim:single:plots:fluid-induced-stresses,fig:sim:single:plots:contact-induced-stresses}, it is evident that the normal stresses exceed the tangential component in magnitude.
After the constriction, the fluid-induced stresses decrease in intensity, while the normal stresses caused by the contacts remain at a comparable level.
In addition, an increase in the tangential stresses caused by contact is observed, which can be attributed to the higher relative velocities involved.
\par
\cref{tab:damage-potential-0.159-30,tab:damage-potential-0.2-30,tab:damage-potential-0.159-60,tab:damage-potential-0.2-60} show the damage potential of the fluid-induced normal (FIN), fluid-induced tangential (FIT), contact-induced normal (CIN), contact-induced tangential (CIT) stresses for $\Rey~\in~\{2,3,5,8\}$, $\alpha~\in~\{0, \frac{\pi}{8}, \frac{\pi}{4}, \frac{3\pi}{8}, \frac{\pi}{2}\}$, expressed in radians, $\Pi_7~\in~\{0.159,0.2\}$, $E~=~30~\si{\kilo\pascal}$, and $E~=~60~\si{\kilo\pascal}$.

\begin{minipage}{\textwidth}
\begin{table}[H]
\caption{Damage potential of FIN, FIT, CIN, CIT stress for $\Pi_7 = 0.159$ and $E = 30~\si{\kilo\pascal}$.\label{tab:damage-potential-0.159-30}}
\centering
\begin{tabular}{llllllllll}
\hline
                       & \multicolumn{4}{c}{{$\Rey=2$}}        &           & \multicolumn{4}{c}{{$\Rey=3$}}        \\ \hline
$\boldsymbol{\alpha}$         & \textbf{FIN} & \textbf{FIT} & \textbf{CIN} & \textbf{CIT} & \textbf{} & \textbf{FIN} & \textbf{FIT} & \textbf{CIN} & \textbf{CIT} \\ \hline
\textbf{$\ang{0}$}    & 0.807       & 0.489       & 0.592       & 0.081       &           & 0.848       & 0.522       & 0.605       & 0.070       \\
\textbf{$\ang{22.5}$} & 0.800       & 0.485       & 0.595       & 0.076       &           & 0.853       & 0.523       & 0.606       & 0.071       \\
\textbf{$\ang{45}$}   & 0.763       & 0.471       & 0.590       & 0.079       &           & 0.838       & 0.510       & 0.598       & 0.073       \\
\textbf{$\ang{67.5}$} & 0.680       & 0.442       & 0.585       & 0.078       &           & 0.783       & 0.481       & 0.599       & 0.071       \\
\textbf{$\ang{90}$}   & 0.453       & 0.210       & 0.574       & 0.098       &           & 0.474       & 0.205       & 0.582       & 0.080       \\ \hline
                       &             &             &             &             &           &             &             &             &             \\ \hline
                       & \multicolumn{4}{c}{{$\Rey=5$}}        &           & \multicolumn{4}{c}{{$\Rey=8$}}        \\ \hline
$\boldsymbol{\alpha}$         & \textbf{FIN} & \textbf{FIT} & \textbf{CIN} & \textbf{CIT} & \textbf{} & \textbf{FIN} & \textbf{FIT} & \textbf{CIN} & \textbf{CIT} \\ \hline
\textbf{$\ang{0}$}    & 0.878       & 0.553       & 0.626       & 0.062       &           & 0.905       & 0.572       & 0.658       & 0.058       \\
\textbf{$\ang{22.5}$} & 0.882       & 0.554       & 0.632       & 0.068       &           & 0.904       & 0.574       & 0.652       & 0.068       \\
\textbf{$\ang{45}$}   & 0.877       & 0.551       & 0.621       & 0.062       &           & 0.905       & 0.576       & 0.649       & 0.066       \\
\textbf{$\ang{67.5}$} & 0.866       & 0.542       & 0.621       & 0.059       &           & 0.908       & 0.577       & 0.661       & 0.063       \\
\textbf{$\ang{90}$}   & 0.489       & 0.239       & 0.592       & 0.061       &           & 0.501       & 0.277       & 0.605       & 0.052       \\ \hline
\end{tabular}
\end{table}
\end{minipage}

\begin{minipage}{\textwidth}
\begin{table}[H]
\caption{Damage potential of FIN, FIT, CIN, CIT stress for $\Pi_7 = 0.2$ and $E = 30~\si{\kilo\pascal}$.\label{tab:damage-potential-0.2-30}}
\centering
\begin{tabular}{llllllllll}
\hline
                       & \multicolumn{4}{c}{{$\Rey=2$}}        &           & \multicolumn{4}{c}{{$\Rey=3$}}        \\ \hline
$\boldsymbol{\alpha}$         & \textbf{FIN} & \textbf{FIT} & \textbf{CIN} & \textbf{CIT} & \textbf{} & \textbf{FIN} & \textbf{FIT} & \textbf{CIN} & \textbf{CIT} \\ \hline
\textbf{$\ang{0}$}    & 0.807       & 0.482       & 0.618       & 0.063       &           & 0.842       & 0.513       & 0.642       & 0.059       \\
\textbf{$\ang{22.5}$} & 0.802       & 0.481       & 0.607       & 0.060       &           & 0.858       & 0.520       & 0.619       & 0.063       \\
\textbf{$\ang{45}$}   & 0.778       & 0.474       & 0.608       & 0.063       &           & 0.846       & 0.511       & 0.622       & 0.056       \\
\textbf{$\ang{67.5}$} & 0.667       & 0.432       & 0.603       & 0.071       &           & 0.794       & 0.485       & 0.623       & 0.060       \\
\textbf{$\ang{90}$}   & 0.452       & 0.167       & 0.585       & 0.093       &           & 0.491       & 0.178       & 0.587       & 0.074       \\ \hline
                       &             &             &             &             &           &             &             &             &             \\ \hline
                       & \multicolumn{4}{c}{{$\Rey=5$}}        &           & \multicolumn{4}{c}{{$\Rey=8$}}        \\ \hline
$\boldsymbol{\alpha}$         & \textbf{FIN} & \textbf{FIT} & \textbf{CIN} & \textbf{CIT} & \textbf{} & \textbf{FIN} & \textbf{FIT} & \textbf{CIN} & \textbf{CIT} \\ \hline
\textbf{$\ang{0}$}    & 0.880       & 0.546       & 0.657       & 0.065       &           & 0.905       & 0.566       & 0.661       & 0.063       \\
\textbf{$\ang{22.5}$} & 0.881       & 0.549       & 0.656       & 0.068       &           & 0.906       & 0.568       & 0.666       & 0.071       \\
\textbf{$\ang{45}$}   & 0.879       & 0.547       & 0.659       & 0.060       &           & 0.907       & 0.569       & 0.666       & 0.057       \\
\textbf{$\ang{67.5}$} & 0.871       & 0.545       & 0.656       & 0.056       &           & 0.907       & 0.567       & 0.685       & 0.062       \\
\textbf{$\ang{90}$}   & 0.496       & 0.216       & 0.592       & 0.058       &           & 0.499       & 0.256       & 0.608       & 0.055       \\ \hline
\end{tabular}
\end{table}
\end{minipage}

\begin{minipage}{\textwidth}
\begin{table}[H]
\caption{Damage potential of FIN, FIT, CIN, CIT stress for $\Pi_7 = 0.159$ and $E = 60~\si{\kilo\pascal}$.\label{tab:damage-potential-0.159-60}}
\centering
\begin{tabular}{llllllllll}
\hline
                       & \multicolumn{4}{c}{{$\Rey=2$}}        &           & \multicolumn{4}{c}{{$\Rey=3$}}        \\ \hline
$\boldsymbol{\alpha}$         & \textbf{FIN} & \textbf{FIT} & \textbf{CIN} & \textbf{CIT} & \textbf{} & \textbf{FIN} & \textbf{FIT} & \textbf{CIN} & \textbf{CIT} \\ \hline
\textbf{$\ang{0}$}    & 0.746       & 0.447       & 0.613       & 0.147       &           & 0.850       & 0.521       & 0.621       & 0.146       \\
\textbf{$\ang{22.5}$} & 0.714       & 0.440       & 0.614       & 0.149       &           & 0.838       & 0.510       & 0.622       & 0.145       \\
\textbf{$\ang{45}$}   & 0.669       & 0.426       & 0.614       & 0.153       &           & 0.840       & 0.512       & 0.623       & 0.152       \\
\textbf{$\ang{67.5}$} & 0.558       & 0.389       & 0.612       & 0.157       &           & 0.783       & 0.476       & 0.623       & 0.146       \\
\textbf{$\ang{90}$}   & 0.707       & 0.319       & 0.621       & 0.148       &           & 0.758       & 0.366       & 0.629       & 0.149       \\ \hline
                       &             &             &             &             &           &             &             &             &             \\ \hline
                       & \multicolumn{4}{c}{{$\Rey=5$}}        &           & \multicolumn{4}{c}{{$\Rey=8$}}        \\ \hline
$\boldsymbol{\alpha}$         & \textbf{FIN} & \textbf{FIT} & \textbf{CIN} & \textbf{CIT} & \textbf{} & \textbf{FIN} & \textbf{FIT} & \textbf{CIN} & \textbf{CIT} \\ \hline
\textbf{$\ang{0}$}    & 0.885       & 0.550       & 0.643       & 0.152       &           & 0.915       & 0.579       & 0.684       & 0.139       \\
\textbf{$\ang{22.5}$} & 0.885       & 0.554       & 0.643       & 0.155       &           & 0.916       & 0.579       & 0.696       & 0.137       \\
\textbf{$\ang{45}$}   & 0.885       & 0.555       & 0.638       & 0.153       &           & 0.917       & 0.578       & 0.685       & 0.140       \\
\textbf{$\ang{67.5}$} & 0.872       & 0.549       & 0.631       & 0.142       &           & 0.912       & 0.577       & 0.680       & 0.137       \\
\textbf{$\ang{90}$}   & 0.780       & 0.401       & 0.652       & 0.153       &           & 0.782       & 0.434       & 0.674       & 0.144       \\ \hline
\end{tabular}
\end{table}
\end{minipage}

\begin{minipage}{\textwidth}
\begin{table}[H]
\caption{Damage potential of FIN, FIT, CIN, CIT stress for $\Pi_7 = 0.2$ and $E = 60~\si{\kilo\pascal}$.\label{tab:damage-potential-0.2-60}}
\centering
\begin{tabular}{llllllllll}
\hline
                       & \multicolumn{4}{c}{{$\Rey=2$}}        &           & \multicolumn{4}{c}{{$\Rey=3$}}        \\ \hline
$\boldsymbol{\alpha}$         & \textbf{FIN} & \textbf{FIT} & \textbf{CIN} & \textbf{CIT} & \textbf{} & \textbf{FIN} & \textbf{FIT} & \textbf{CIN} & \textbf{CIT} \\ \hline
\textbf{$\ang{0}$}    & 0.839       & 0.507       & 0.647       & 0.156       &           & 0.869       & 0.531       & 0.648       & 0.157       \\
\textbf{$\ang{22.5}$} & 0.827       & 0.501       & 0.646       & 0.156       &           & 0.873       & 0.533       & 0.652       & 0.161       \\
\textbf{$\ang{45}$}   & 0.803       & 0.488       & 0.639       & 0.163       &           & 0.871       & 0.531       & 0.651       & 0.163       \\
\textbf{$\ang{67.5}$} & 0.680       & 0.430       & 0.632       & 0.153       &           & 0.817       & 0.496       & 0.646       & 0.148       \\
\textbf{$\ang{90}$}   & 0.689       & 0.295       & 0.641       & 0.173       &           & 0.723       & 0.329       & 0.649       & 0.167       \\ \hline
                       &             &             &             &             &           &             &             &             &             \\ \hline
                       & \multicolumn{4}{c}{{$\Rey=5$}}        &           & \multicolumn{4}{c}{{$\Rey=8$}}        \\ \hline
$\boldsymbol{\alpha}$         & \textbf{FIN} & \textbf{FIT} & \textbf{CIN} & \textbf{CIT} & \textbf{} & \textbf{FIN} & \textbf{FIT} & \textbf{CIN} & \textbf{CIT} \\ \hline
\textbf{$\ang{0}$}    & 0.878       & 0.541       & 0.674       & 0.150       &           & 0.914       & 0.570       & 0.726       & 0.141       \\
\textbf{$\ang{22.5}$} & 0.883       & 0.547       & 0.667       & 0.157       &           & 0.917       & 0.573       & 0.716       & 0.145       \\
\textbf{$\ang{45}$}   & 0.886       & 0.551       & 0.663       & 0.161       &           & 0.917       & 0.570       & 0.718       & 0.146       \\
\textbf{$\ang{67.5}$} & 0.878       & 0.550       & 0.656       & 0.158       &           & 0.913       & 0.572       & 0.715       & 0.138       \\
\textbf{$\ang{90}$}   & 0.734       & 0.364       & 0.672       & 0.159       &           & 0.753       & 0.410       & 0.692       & 0.154       \\ \hline
\end{tabular}
\end{table}
\end{minipage}
\newpage

In general, the damage potential of FIN stresses is the highest, followed by CIN stresses and FIT stresses.
All these stresses are on a similar relatively high level, only the CIT stresses have a minor damage potential.
\par
Furthermore, the damage potential for all stresses increases with $\Rey$.
This trend is consistent across all the tables and angles.
The increase in damage potential with $\Rey$ is more pronounced for FIN and FIT stresses than for CIN and CIT stresses.
\par
The damage potential for FIN, FIT, CIN, and CIT stresses also varies with the angle $\alpha$.
For $\Rey\leq3$, the angle, especially for $\alpha \geq \ang{67.5}$, has higher influence than for $\Rey\geq5$.
However, in across all parameters, a big difference in FIN and FIT stresses is observed for $\alpha=\ang{90}$, i.e. a pipe without any cross-section constriction.
This is also observed for the damage potential due to CIN stresses, but only for $E = 30~\si{\kilo\pascal}$, for $E = 60~\si{\kilo\pascal}$ those differences vanish.
\par
The changes to the damage potential for FIN and FIT stresses are only small for the considered sizes $\Pi_7~=~0.159$ and $\Pi_7~=~0.2$.
However, especially at $E = 60~\si{\kilo\pascal}$, the CIN and CIT stresses cause a higher damage potential for the bigger particles.
\par
For FIN and FIT stresses the damage potential varies with the Young's modulus $E$.
However, there is no consistent trend.
The damage potential due to CIN and CIT stresses, on the other hand, increases consistently with $E$ and the impact is also pronounced.
\par
In general, the biggest damage potentials result for $Re~=~8$, $\alpha~<~\ang{90}$, $\Pi_7~=~0.2$, and $E = 60~\si{\kilo\pascal}$, see \cref{tab:damage-potential-0.2-60}.
\par
The strong dependence on the Reynolds number $\Rey$ is due to the fact that $\Rey$ represents the ratio of inertial forces to viscous forces in a fluid.
For fluid-induced stresses (FIN and FIT), the increased inertial forces at higher Reynolds numbers cause the fluid to exert more force on the surrounding surfaces.
Also, the velocity gradients within the fluid become higher, resulting in higher shear stresses.
In the context of contact-induced stresses (CIN and CIT), the increased inertial forces at higher Reynolds numbers result in more forceful solid-solid impacts, resulting in higher damage potential.
\par
Compared to the impact of $\Rey$, the angle $\alpha$ has only a small impact, which is only noteworthy for smaller $\Rey$.
However, in these cases, it is evident that a higher taper angle reduces the damage potential, because the more sudden the change in the diameter, the higher the chance for contacts and the bigger the changes in the velocity gradient.
\par
The particle properties, Young's modulus and particle size, influence mainly the conduct-induced stresses (CIN and CIT).
This is because the contact forces directly depend on the Young's modulus and higher particle sizes of the as rigid considered particles make contacts more likely.
\par
In summary, the above observations highlight that $\Rey$ has the biggest impact on the damage potential.
The influence of $\alpha$ is only measurable for $\Rey\leq3$ and otherwise negligible.
In the measurable cases, a larger taper angle leads to a generally smaller damage potential, for $\alpha~\geq~\ang{45}$.
The strong change for $\alpha~=~\ang{90}$ is actually caused by the change of $\Rey$ in the cross-section constriction.
Furthermore, the damage potential due to contact depends strongly on the mechanical properties and size of the particle.

\subsection{Validation}
\label{sec:results:validation}

The underlying computational framework has already been rigorously validated~\cite{Krause_Klemens_Henn_Trunk_Nirschl_2017,Trunk_Marquardt_Thaeter_Nirschl_Krause_2018,Trunk_Weckerle_Hafen_Thaeter_Nirschl_Krause_2021,Marquardt2024a,Marquardt_Römer_Nirschl_Krause_2023,Marquardt2024b} and the velocity profiles obtained from PIV measurements on Herschel--Bulkley fluids could be reproduced by the simulations~\cite{eysel2023}.
This section extends these validations by comparing the experimental results in Section~\ref{sec:results:exp} with the computational results in Section~\ref{sec:results:comp}, focusing on the key observations that emerged from both approaches.
\par
There is a strong correlation between the experimental and computational results.
First, both methods revealed a significant dependence of the damage potential on the Reynolds number.
As the Reynolds number increases, we consistently observed a corresponding increase in the damage potential in both experimental and numerical investigations.
Second, our experimental and computational approaches agreed on the relatively small effect of geometry, specifically the taper angle, on the damage potential, as the damage potential due to the angle is much less pronounced compared to the dominant influence of the Reynolds number.
\par
The agreement between experimental and numerical results in these key areas serves as a validation of our research methodology and results. This agreement not only reinforces the reliability of each approach, but also provides a more comprehensive and confident understanding of the phenomena under study. The consistency of the observations suggests that our numerical models accurately capture the essential physics of the system, lending credibility to both the experimental setup and the computational framework.

\subsection{Discussion}
\label{sec:results:discussion}

Our findings, derived from both numerical and experimental studies, consistently point to the local Metzner--Reed Reynolds number as the most important parameter influencing the damage potential of particles during transport.
This insight is critical because the Reynolds number is influenced by both process and design parameters, providing multiple avenues for reducing damage potential.
In practice, there are several options, such as adjusting the flow rate or modifying the geometry to achieve optimal conditions.
\par
Our research suggests that fluid deflection may not be a significant problem, primarily due to the nature of the fluid being considered, since the impact of the taper angle is less pronounced than that of the Reynolds number.
However, our results indicate that it can still influence damage potential under certain conditions.
This effect is particularly noticeable when the taper angle changes the local Reynolds number, highlighting the interrelated nature of these parameters.
In addition, particle-specific factors such as Young's modulus and size have been found to influence damage potential.
However, it is important to note that these parameters are largely dependent on the type of fruit and its growing conditions, making them less controllable in practical applications.
\par
The importance of the Metzner--Reed Reynolds number could also be applied to other critical components in transportation systems, such as valves and manifolds.
Our results suggest that maintaining an approximately constant Reynolds number also in these parts could be an effective strategy for achieving uniform damage potential throughout the transport system.
This approach could lead to more consistent design principles for different components of particle transport systems, particularly in industries that handle delicate materials such as fruit.
This could lead to reduced waste, improved product quality, and potentially economic benefits in industries where minimizing particle damage is critical.
\par
It is important to emphasize that while our results strongly suggest these conclusions, they require further validation across more possible scenarios and geometries.
Applying our results to other geometries, such as aseptic valves, is a promising direction for future research, since these complex geometries could not be investigated in this study.
Further studies are needed to validate the broader applicability of maintaining constant Reynolds numbers as a damage reduction strategy for various system components.
\par
Besides investigating more system components, future research could also explore the applicability of these findings to a wider range of particle types and fluid properties. However, the experiments with two different fluids showed similar damage potential at comparable Reynolds numbers.
In addition, long-term studies in real-world applications would be valuable to confirm the practical benefits of designing systems based on Reynolds number optimization.
\par
In conclusion, our study provides a basis for understanding and potentially controlling particle damage in transport systems.
By emphasizing the importance of the Reynolds number and providing insights into the role of geometry and particle properties, the results highlight potential optimizations in system design.
In addition, these results emphasize the need for further, more in-depth studies considering a wider range of materials and system components to determine further potential improvements in real-world systems.

\section{Summary and conclusions}
\label{sec:summary}

This study investigates the potential for damage of fruits immersed in a Herschel--Bulkley fluid during transport through cross-section constrictions, using both experiments and simulations.
Industrial peach preparations are therefore conveyed in three different pipe geometries at a technical scale. Two different volume flow rates are applied, and the disintegration of the fruit pieces is quantified by wet sieving.  
Four-way coupled fully resolved particulate flow simulations are realized using \HLBM{}.
Its goal is the quantification of the damage potential and also the identification of parameters with a strong correlation.
\par
The results reveal that the local Reynolds number has the most significant impact on the damage potential.
The taper angle, a geometric parameter, has a limited effect, but it can influence the damage potential, especially if it influences the local Reynolds number.
The particle parameters, including Young's modulus and size, also contribute to the damage potential, but they are largely dependent on the type of fruit and its growing conditions.
\par
The findings therefore suggest that the damage potential is reducible by adjusting the process conditions or by modifying the geometry to control the local Reynolds number.
This, in turn, results in end products of higher quality and value.

\textbf{\emph{Acknowledgements:}}
This IGF Project of the FEI is/was supported within the programme for promoting the Industrial Collective Research (IGF) of the Federal Ministry of Economic Affairs and Climate Action (BMWK), based on a resolution of the German Parliament. Project 21096 N.
This work was performed on the HoreKa supercomputer funded by the Ministry of Science, Research and the Arts Baden-Württemberg and by the Federal Ministry of Education and Research.
The authors acknowledge support by the state of Baden-Württemberg through bwHPC.
The authors of the Technical University of Berlin would like to thank for the funding of Deutsche Forschungsgemeinschaft (DFG) for the multiscale Particle-Image-Velocimetry (PIV)-system in the frame of project 431729792.

\textbf{\emph{Author contribution statement:}}
\textbf{J.\ E.\ Marquardt}: Conceptualization, Methodology, Software, Validation, Formal analysis, Investigation, Data curation, Writing - Original Draft, Writing - Review \& Editing, Visualization, Project administration, Funding acquisition; 
\textbf{B.\ Eysel}: Conceptualization, Methodology, Validation, Formal analysis, Investigation, Data curation, Writing - Original Draft, Writing - Review \& Editing, Visualization, Project administration.
\textbf{M.\ Sadric}: Software, Formal analysis, Data curation, Writing - Review \& Editing.
\textbf{C.\ Rauh}: Resources, Supervision, Project administration, Funding acquisition.
\textbf{M.\ J.\ Krause}: Software, Resources, Supervision, Project administration, Funding acquisition.


\end{document}